\documentclass[preprint,3p]{elsarticle}

\usepackage{amssymb}
\usepackage[version=3]{mhchem}
\usepackage{hyperref}
\usepackage{graphicx}
\usepackage{subcaption}
\usepackage{bm}
\usepackage{multirow}
\usepackage{makecell}

\journal{Nuclear Engineering and Technology}

\begin{document}

\begin{frontmatter}



\title{Towards energy-insensitive and robust neutron/gamma classification: A learning-based frequency-domain parametric approach}


\author[myfirstaff,mysecondaff]{Pengcheng Ai\corref{mycorrespondingauthor}}
\ead{aipc@ccnu.edu.cn}

\author[myfirstaff,mysecondaff]{Hongtao Qin}

\author[myfirstaff,mysecondaff]{Xiangming Sun}

\author[myfirstaff,mysecondaff]{Kaiwen Shang}

\cortext[mycorrespondingauthor]{Corresponding author}

\affiliation[myfirstaff]{organization={PLAC, Key Laboratory of Quark and Lepton Physics (MOE), Central China Normal University},
            addressline={No. 152 Luoyu Road}, 
            city={Wuhan},
            postcode={430079}, 
            state={Hubei},
            country={China}}

\affiliation[mysecondaff]{organization={Hubei Provincial Engineering Research Center of Silicon Pixel Chip \& Detection Technology},
	addressline={No. 152 Luoyu Road}, 
	city={Wuhan},
	postcode={430079}, 
	state={Hubei},
	country={China}}

\begin{abstract}
Neutron/gamma discrimination has been intensively researched in recent years, due to its unique scientific value and widespread applications. With the advancement of detection materials and algorithms, nowadays we can achieve fairly good discrimination. However, further improvements rely on better utilization of detector raw signals, especially energy-independent pulse characteristics. We begin by discussing why figure-of-merit (FoM) is not a comprehensive criterion for high-precision neutron/gamma discriminators, and proposing a new evaluation method based on adversarial sampling. Inspired by frequency-domain analysis in existing literature, parametric linear/nonlinear models with minimum complexity are created, upon the discrete spectrum, with tunable parameters just as neural networks. We train the models on an open-source neutron/gamma dataset (CLYC crystals with silicon photomultipliers) preprocessed by charge normalization to discover and exploit energy-independent features. The performance is evaluated on different sampling rates and noise levels, in comparison with the frequency classification index and conventional methods. The frequency-domain parametric models show higher accuracy and better adaptability to variations of data integrity than other discriminators. The proposed method is also promising for online inference on economical hardware and portable devices.
\end{abstract}


\end{frontmatter}


\section{Introduction}

Neutron detection is a classic problem researched intensively in nuclear societies, featuring an increasing popularity in various applications from physics experiments to radiation protection. The problem is most challenging owing to the mixture of neutron and gamma signals both detected by the sensitive volume, with only slight differences in pulse shapes. Neutron/gamma discrimination based on pulse shape discrimination (PSD) aims to separate neutron and gamma events by waveform sampling and dedicated analysis algorithms. Both inorganic and organic scintillators \cite{Ahnouz01122024}, coupled to photomultipliers \cite{SODERSTROM2019238} or silicon photomultipliers (SiPM) \cite{LEE2024169638}, are usable for neutron detection, while inorganic scintillators have higher sensitivity and organic ones are more cost-effective.

In recent years, algorithms for neutron/gamma discrimination have been improved greatly. While conventional time-domain analog algorithms \cite{WOLSKI1995584,BAYAT2012217} have laid a solid foundation and put into use in many applications, latest algorithms focus more on frequency-domain analysis \cite{5485131,Liu_2016} and digital signal processing \cite{YE2022166256,Liu2021,Liu2022,LIU20233359} after analog-to-digital converters (ADC). Furthermore, influenced by the artificial intelligence breakthrough, machine learning (ML) methods \cite{FABIAN2021164750,YOON20233925,Zhao_2023,DOUCET2020161201,HACHEM20234057} show competitive performance in the area of neutron/gamma discrimination, extending discrimination power to low energy regions \cite{ZHANG2024111179,10672537} and pile-ups \cite{PAN2024103329,HAN2022166328,Peng_2022}. With the state-of-the-art algorithms and under proper experimental conditions, reaching figure-of-merit (FoM) above 1.2 and accuracy above 98\% would be common cases for practical uses.

Considering the achieved progress, it is intriguing to contemplate the future direction of performance optimization on neutron/gamma discrimination. Although FoM is a widely used evaluation criterion for its simplicity, universality and unsupervised nature (without event labelling), the curve fitting procedure tends to be biased by majorities, and the goodness of fit is not covered by the single FoM number. This limitation becomes much more pronounced when the discrimination power becomes stronger. Besides, for specific datasets, a considerable amount of discrimination ability comes from the energy gap between neutron and gamma events, which may hinder further improvements, especially for exploitation-dominated ML methods. These factors motivate us to develop improved algorithms insensitive to energy differences, under a more reasonable criterion for performance on critical events.

In this paper, we propose a new approach to classify neutron/gamma events based on frequency-domain parametric models (FDPM). This approach is partially inspired by \cite{MORALES2024745}, which uses frequency classification index as the discrimination indicator to separate two types of events. Our approach is validated on an open-source CLYC scintillator dataset \cite{CLYCDataset} (also used by \cite{MORALES2024745}). CLYC (\ce{Cs2LiYCl6}:\ce{Ce}) is an emerging scintillator capable of detecting thermal neutrons, gamma rays and fast neutrons, and used for neutron/gamma PSD in the recent literature \cite{Zhao_2023,DOUCET2020161201,HAN2022166328}. The major contributions of the paper include:

\begin{itemize}
	\item A new neutron/gamma classifier base on FDPM, in its typical form of simple neural networks, for efficient and effective neutron/gamma event discrimination.
	\item A newly proposed evaluation method for high-precision neutron/gamma discriminators, using receiver operating characteristic (ROC) curves, with adversarial sampling from the dataset.
	\item Validation of the proposed classifier with the proposed evaluation method on an open-source CLYC scintillator dataset to display its superior performance.
\end{itemize}

\section{Evaluation criteria}

\subsection{The FoM criterion and its limitation}

\begin{figure}[htb]
	\centering
	\begin{subfigure}{0.32\textwidth}
		\includegraphics[width=\textwidth]{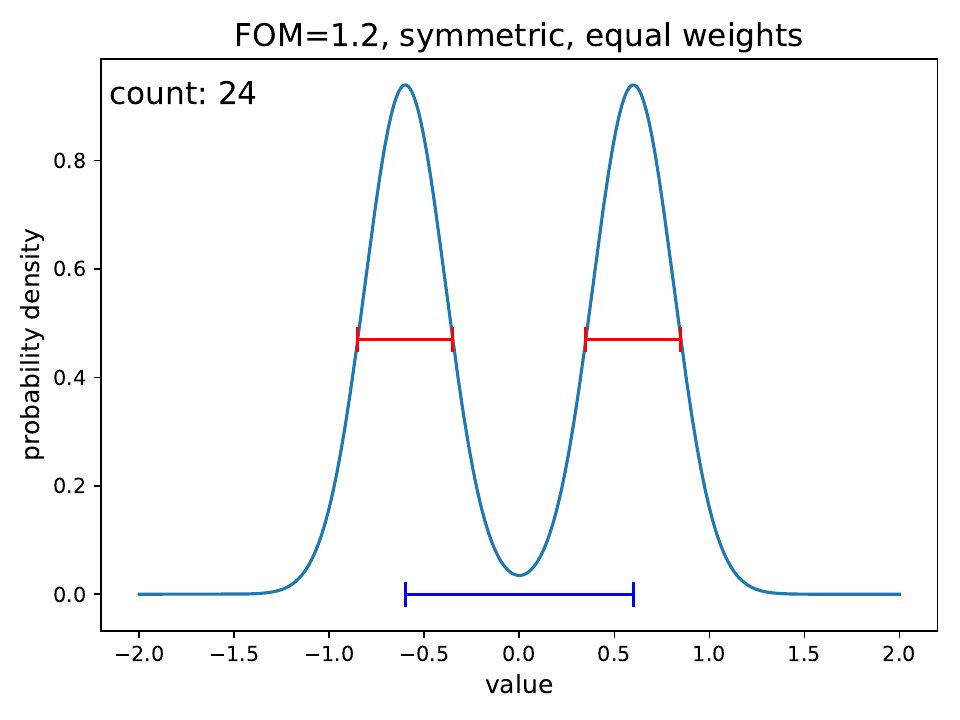}
		\caption{}
		\label{fig:sym-equ-ideal}
	\end{subfigure}
	\hfill
	\begin{subfigure}{0.32\textwidth}
		\includegraphics[width=\textwidth]{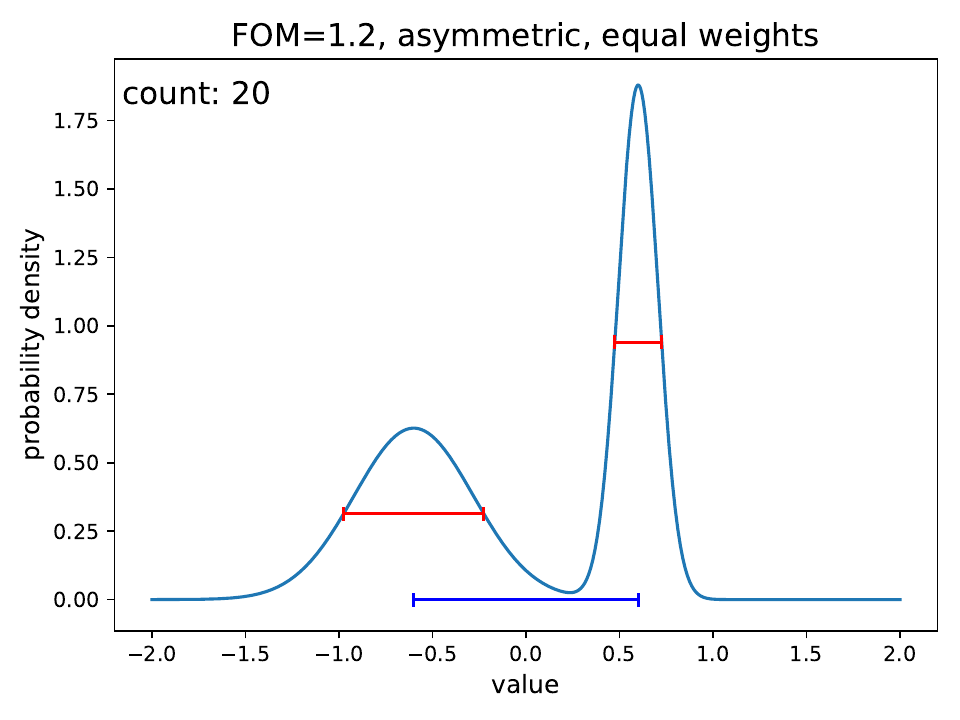}
		\caption{}
		\label{fig:asym-equ-ideal}
	\end{subfigure}
	\hfill
	\begin{subfigure}{0.32\textwidth}
		\includegraphics[width=\textwidth]{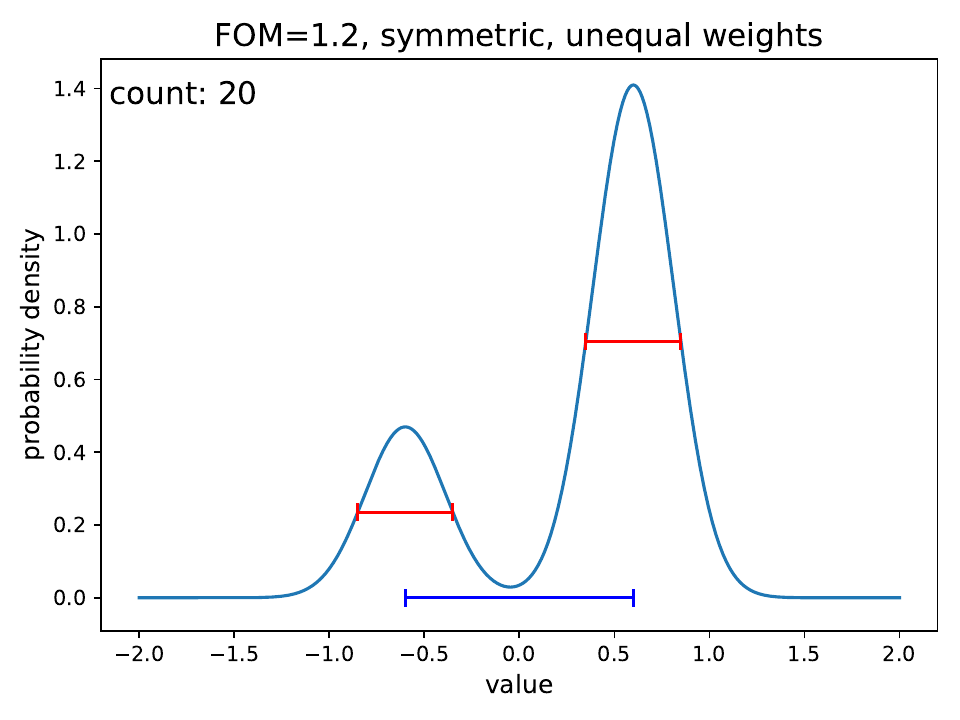}
		\caption{}
		\label{fig:sym-unequ-ideal}
	\end{subfigure}
	\begin{subfigure}{0.32\textwidth}
		\includegraphics[width=\textwidth]{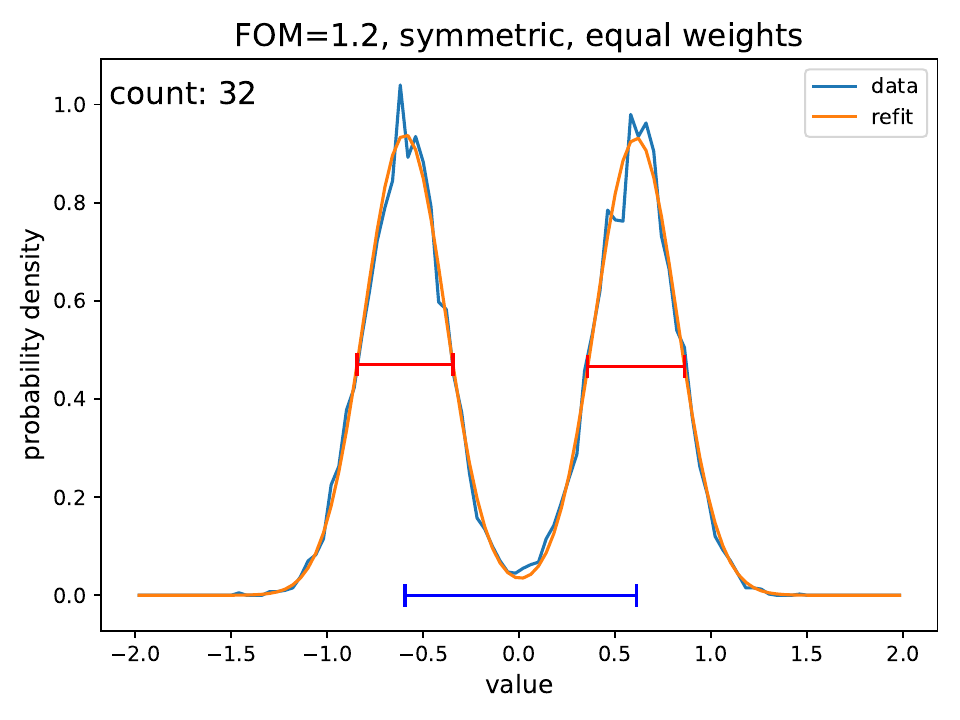}
		\caption{}
		\label{fig:sym-equ-refit}
	\end{subfigure}
	\hfill
	\begin{subfigure}{0.32\textwidth}
		\includegraphics[width=\textwidth]{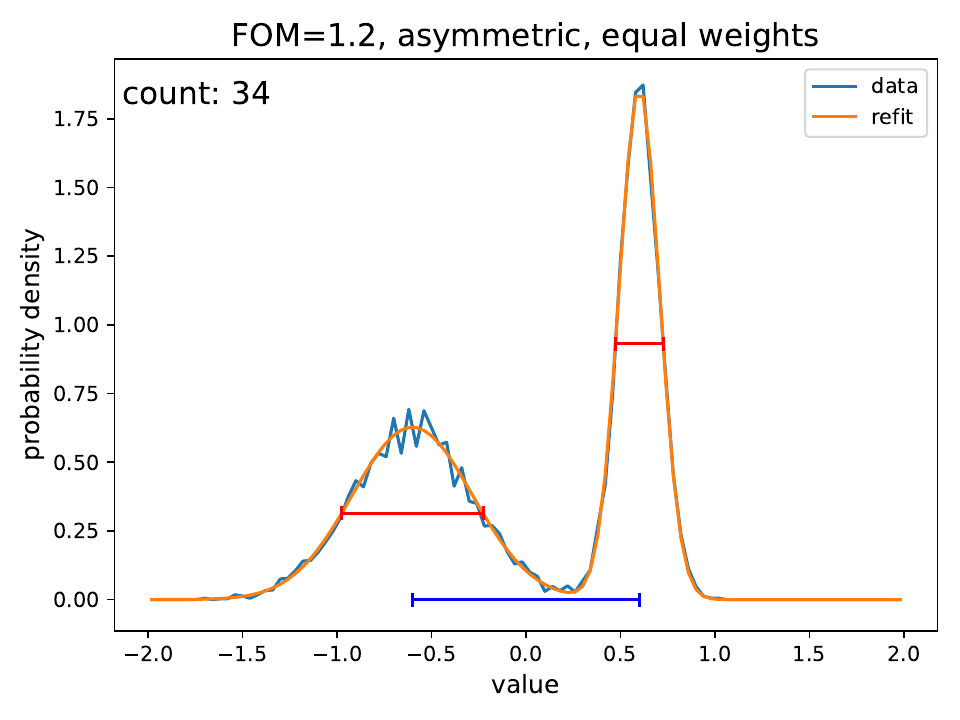}
		\caption{}
		\label{fig:asym-equ-refit}
	\end{subfigure}
	\hfill
	\begin{subfigure}{0.32\textwidth}
		\includegraphics[width=\textwidth]{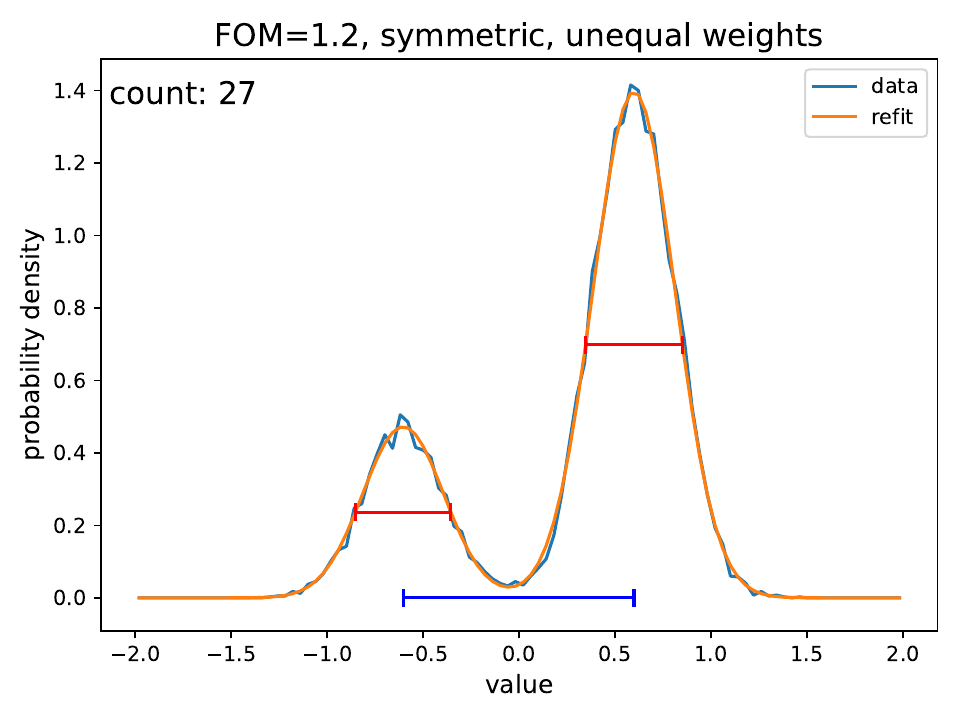}
		\caption{}
		\label{fig:sym-unequ-refit}
	\end{subfigure}
	\caption{These sub-figures illustrate that distributions with different numbers of misclassified examples can share the same FoM. Each figure is generated with 10000 examples in total. The misclassified count is indicated at the top left in each sub-figure. Sub-figures in the bottom row add noise to each example and refit a double-normal distribution to these noisy examples.}
	\label{fig:fom-limit}
\end{figure}

FoM is the most commonly used criterion for neutron/gamma distribution in literature. Given a classification index, it is defined in equation (\ref{equ:fom}):

\begin{equation}\label{equ:fom}
	\mathrm{FoM} = \frac{|\mu_1 - \mu_2|}{\mathrm{FWHM}_1 + \mathrm{FWHM}_2}
\end{equation}

\noindent where $\mu_1, \mu_2$ and $\mathrm{FWHM}_1, \mathrm{FWHM}_2$ stand for the mean values and the full-width half-maxima (FWHM) for the two fitted normal peaks. From its definition, we can find that computing FoM on a classification index is simple and straightforward; we only need to fit a double-normal distribution to the mixture of neutron and gamma events without extra information. Actually, it is quite useful when labelling the ground-truth event for each observed signal is hard or impossible.

However, FoM has its limitation as not being able to characterize details of distributions and critical events near the crossing line, especially for high-precision discrimination. Fig. \ref{fig:fom-limit} gives an explanatory illustration for this fact. In the top row, we generate original double-normal distributions with (a) symmetric standard deviations, equal weights on two components; (b) asymmetric standard deviations, equal weights on two components; and (c) symmetric standard deviations, unequal weights on two components. The classification crossing point is optimally determined by each distribution. It can be seen that although all three distributions have an FoM of 1.2, the misclassified count varies between 20 to 24 per 10000 events, with 20\% fluctuations. In the bottom row, noise has been added to data obeying corresponding distributions in the above sub-figures, and we refit double-normal distributions to these noisy data and ensure FoM as the same. After noise addition, the misclassified count shows a much greater divergence from 27 to 34, which equals to 70\% fluctuations. Therefore, when focusing on the misclassified minorities by high-precision discriminators, FoM is no longer an accurate enough criterion.

Furthermore, computing FoM for ML-based classification indexes usually falls short of its original purpose. Under sufficient optimization, two peaks of the double-normal distribution on the index can be pushed far away from each other; however, misclassified examples can still be observed noticeably, which implies a ``false hood of confidence''. This phenomenon, along with the above limitation, motivates us to find other supplementary criteria for neutron/gamma discrimination.

\subsection{ROC curves with adversarial sampling}
\label{sec:roc-adv}

\begin{figure}[htb]
	\centering
	\begin{subfigure}{0.32\textwidth}
		\includegraphics[width=\textwidth]{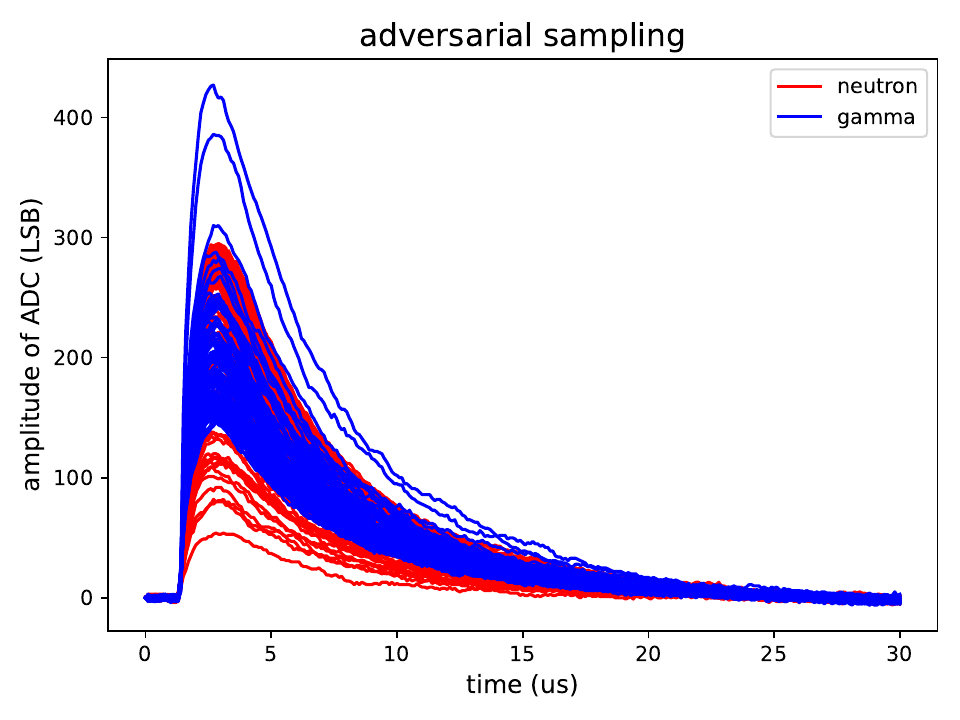}
		\caption{}
		\label{fig:ng-adv}
	\end{subfigure}
	\begin{subfigure}{0.32\textwidth}
		\includegraphics[width=\textwidth]{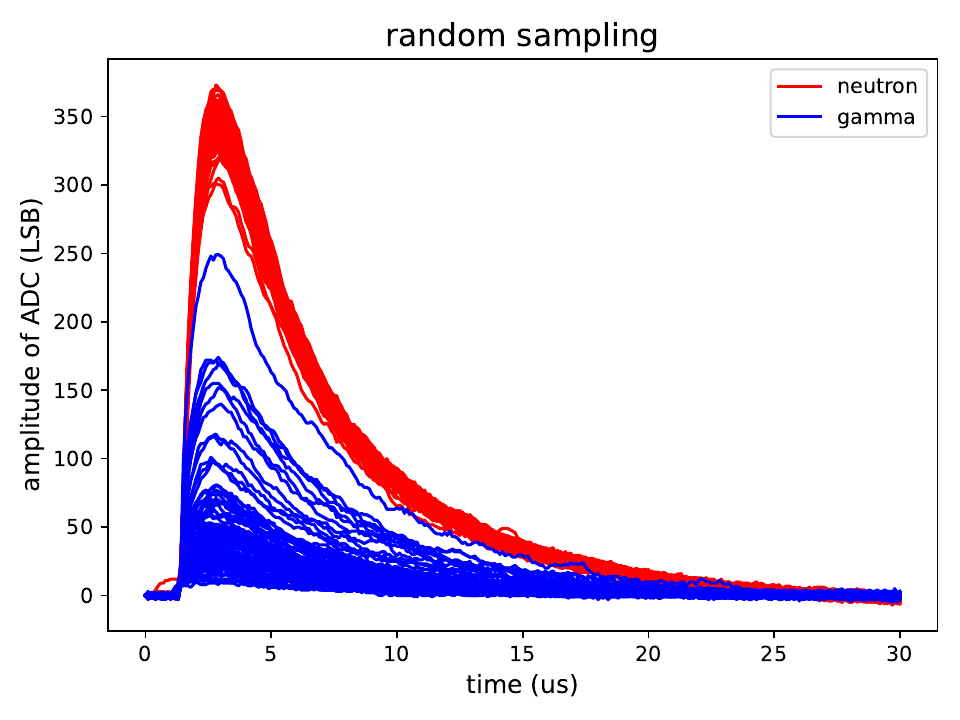}
		\caption{}
		\label{fig:ng-random}
	\end{subfigure}
	\caption{Illustration of 100 neutron examples and 100 gamma examples with (a) adversarial sampling and (b) random sampling.}
	\label{fig:ng-samples}
\end{figure}

Since the target of most neutron/gamma discriminators is to separate neutron events apart from gamma events, we treat neutron events as signals and gamma events as backgrounds. With this setting, the original problem can be reformulated as a binary classification task with an emphasis on background suppression.

Receiver operating characteristic (ROC) is a univariate two-dimensional plot commonly used in information retrieval with binary categories. If we denote signals as 1 and backgrounds as 0, signal efficiency (SE) and background rejection (BR) at a threshold ($t$) of classification index ($I$) can be defined as:
\begin{align}
	\mathrm{SE} &= P(\hat{y}(t) = 1|y = 1) = \frac{P(I_y \geq t, y = 1)}{P(y = 1)} \\
	\mathrm{BR} &= P(\hat{y}(t) = 0|y = 0) = \frac{P(I_y < t, y = 0)}{P(y = 0)}
\end{align}

\noindent where $I_y$ is the classification index associated with the example $y$. It can be seen that both SE and BR are functions of the threshold $t$. When the threshold varies, we can get a curve (with SE as x-axis and BR as y-axis) starting from $(0, 1)$ and ending at $(1, 0)$. The curve stands for the discrimination ability provided by the classification index. It is quite useful to measure the comprehensive ability to separate binary categories even when positive and negative events are unequal.

However, high-precision neutron/gamma discriminators tend to generate ROC curves which are too ideal (like two line segments from $(0, 1)$ to $(1, 1)$ and from $(1, 1)$ to $(1, 0)$) to give useful clues about which one is better, because the curves are dominated by the majorities being trivially well classified. To better examine the classification ability on critical events near the crossing point, adversarial sampling could be used to focus on these critical events. The principle of adversarial sampling is to select examples which are shared by several discrimination methods and critical enough to cover most errors from predictions of these methods. The adversarial sampling strategy for the mainly discussed CLYC dataset is shown in Fig. \ref{fig:ng-samples}. It can be seen that, in normal random sampling, the amplitudes of neutron events are significantly greater, which can be used as an extra handle beyond pulse shapes. When adversarial sampling is used, neutron events with smallest amplitudes and gamma events with largest amplitudes are selected, which have nearly the same energy distribution. The adversarial sampling makes the classification much more challenging, and ROC curves on these adversarial samples can be much more informative. In this paper, we mainly use ROC curves with adversarial sampling as our evaluation criterion. It should be noted that, different data distributions could lead to totally different adversarial sampling strategies (section \ref{sec:add-tof}).

\section{Method}

\begin{figure}[htb]
	\centering
	\includegraphics[width=0.6\textwidth]{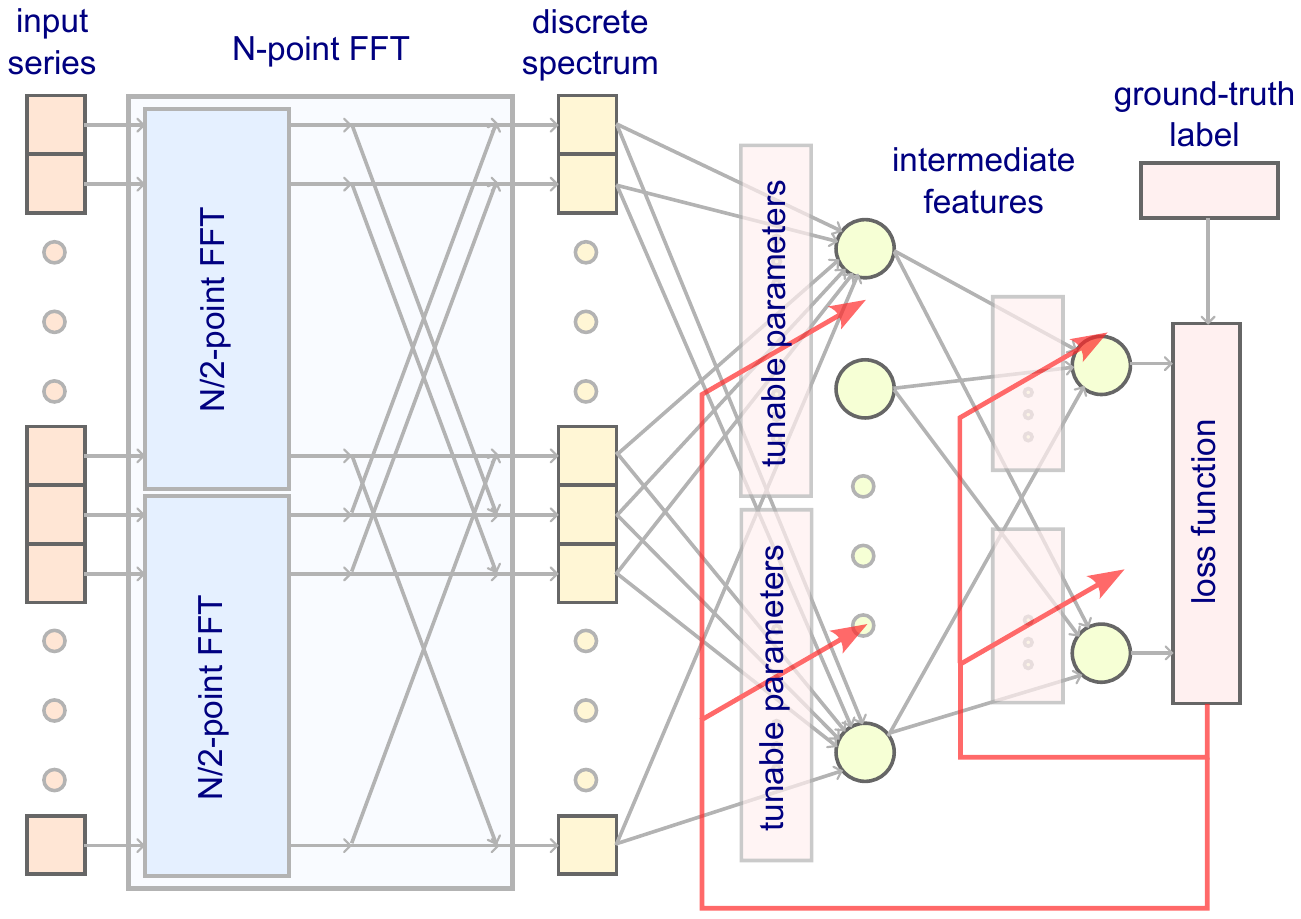}
	\caption{\label{fig:method} Illustration of the proposed FDPM. The parameters are tuned offline and deployed online for inference.}
\end{figure}

Conventional methods for neutron/gamma discrimination generate the classification index with a fixed procedure, such as charge comparison or frequency gradient analysis. In \cite{MORALES2024745}, the authors use Fast Fourier Transform (FFT) on raw waveforms, and use the ratio between the low-frequency portion and the wide-frequency portion on the spectrum as the classification index. This method achieves considerably good results on a CLYC scintillator dataset collected also by the authors.

Inspired by the above method, we propose an FDPM based on FFT and simple feedforward neural networks, as shown in Fig. \ref{fig:method}. Instead of using fixed boundaries for low-frequency and wide-frequency portions, this model tries to learn parameters related to the full FFT spectrum for optimal classification and feature extraction. Thanks to the high-quality labelled dataset provided in \cite{CLYCDataset}, the model can be trained in a supervised manner, and parameters can be tuned with gradient descent and back-propagation.

In principle, any parametric model can be built upon the FFT spectrum, as long as its parameters can be effectively adjusted. In this paper, however, we focus on simple feedforward models because of their simplicity and interpretability. Specifically, we investigate two variants of feedforward models, the linear model and the nonlinear model. Their mathematical forms can be written as:
\begin{align}
	\bm{f}^c &= \mathrm{FFT}(\bm{s}),\quad \bm{s} \in \mathbb{R}^N, \bm{f}^c \in \mathbb{C}^N, \label{equ:fft} \\
	\bm{f}^r &= |\mathrm{Re}(\bm{f}^c)| + |\mathrm{Im}(\bm{f}^c)|,\quad \bm{f}^r \in \mathbb{R}^N, \label{equ:cb} \\
	\bm{p}^l &= \bm{W}^l \bm{f}^r + \bm{b}^l,\quad \bm{W}^l \in \mathbb{R}^{2 \times N}, \bm{b}^l \in \mathbb{R}^2, \label{equ:linear} \\
	\bm{p}^n &= \bm{W}_2^n \sigma (\bm{W}_1^n \bm{f}^r + \bm{b}_1^n) + \bm{b}_2^n,\quad \bm{W}_1^n \in \mathbb{R}^{\frac{N}{2} \times N}, \bm{b}_1^n \in \mathbb{R}^{\frac{N}{2}}, \bm{W}_2^n \in \mathbb{R}^{2 \times \frac{N}{2}}, \bm{b}_2^n \in \mathbb{R}^2. \label{equ:nonlinear}
\end{align}

\noindent where $\bm{s}$ stands for the input time series, $\bm{W}/\bm{b}$ stands for tunable parameters, and $\bm{p}^l$/$\bm{p}^n$ stands for the output of the linear/nonlinear model, respectively. The function $\sigma(\cdot)$ stands for an element-wise nonlinear activation function, which can be the rectified linear unit (ReLU) \cite{DBLP:conf/icml/NairH10} or the sigmoid function. In equation (\ref{equ:fft}), a full discrete frequency spectrum (with the same length as input) is computed. In equation (\ref{equ:cb}), we use the city-block approach \cite{ALHARBI2019205} to convert complex numbers to real numbers, while other methods (such as taking the absolute value) can also be used. The main reason for this approach is its simplicity for computation especially for embedded devices; besides, this is in line with the frequency-domain method \cite{MORALES2024745} we compare against for fair comparison. In equation (\ref{equ:linear}), a weight matrix, with shape $2 \times N$, maps the discrete frequency to two targets (neutron and gamma), and a bias vector, with length of 2, is added to the result of matrix-vector multiplication. the larger value of the two targets will be the predicted class. In equation (\ref{equ:nonlinear}), two steps are performed. The first step is to map the discrete frequency ($N$) to half its length ($N/2$), with a bias vector being added. The reason to choose $N/2$ is to provide sufficient approximation power, while smaller values can be tried to trade off between accuracy and compute-intensity. Afterwards, a nonlinear activation function is applied. The second step is to map the activated vector ($N/2$) to two targets just as the linear model.

The loss function is computed by the cross entropy:
\begin{equation}
	\mathcal{L} = -\sum\limits_{i=0,1} \log \frac{\exp(p_i)}{\sum_{j=0,1}\exp(p_j)} \cdot \bm{1}\{i = y\}
\end{equation}

\noindent where $p_i(i=0,1)$ stands for the binary output of the linear/nonlinear model, $y$ stands for the ground-truth label corresponding to the example, and $\bm{1}\{\cdot\}$ takes 1 when the condition is satisfied and 0 otherwise. The gradients of the loss function with regard to the parameters are computed, and the values of parameters are adjusted towards the gradient descent direction. The model is trained with the loss function until parameters are sufficiently optimized.

\section{Experimental results}

\subsection{Experiment setup}

We use an open-source neutron/gamma discrimination dataset \cite{CLYCDataset} collected by the CLYC crystal coupled to a SiPM array (OnSemi ArrayC-60035-4P). The detector (CLYC crystal, SiPM, amplifier, bias) is a commercial one \cite{detector-frontend} sealed and produced by Scionix B.V. More characteristics about the detector are discussed in section \ref{sec:add-noise-com}. The raw nuclear pulses are recorded by a 10-bit digitizer (CAEN DT5761) at 4 GS/s, and then downsampled to 100 MS/s. The dataset consists of 27696 neutron events (by deuterium--tritium and deuterium--deuterium generators) and 10913 gamma events (by Cs-137).

To prepare the dataset for training and testing, we divide the original dataset into the training set, validation set and test set with a ratio of 7:1:2. This results in 19387 neutron and 7639 gamma events in the training set, 2769 neutron and 1091 gamma events in the validation set, and 5540 neutron and 2183 gamma events in the test set. To prevent models from learning unintended patterns and converging to trivial results, we perform the following preprocessing steps before sending the raw signal to the models:

\begin{enumerate}
	\item The baseline is subtracted from the signal, and the amplitude is inverted to convert negative pulses to positive pulses.
	\item A random starting point, ranging from 0 to 99, is chosen to resample the signal to eliminate possible mismatches between neutron and gamma events.
	\item (Optional) To research different sampling rates, subsampling is performed with the starting point to generate sampling rates from 100 MS/s to 10 MS/s.
	\item (Optional) To research different noise conditions, different kinds of noise are added to the signal with the specific signal-to-noise ratio (SNR) relative to the maximum signal amplitude.
	\item Finally, to force models to learn energy-independent features from the signal, all examples are normalized to have the same integral values.
\end{enumerate}

It should be noted that the final step is essential for models to be energy-insensitive. This will be shown in the subsection below.

In the following experiments, we train the models on a desktop computer with two NVIDIA RTX 4060 Ti graphic cards (2$\times$16 GB video memory). For linear models, we use the stochastic gradient descent with momentum for optimization. For nonlinear models, we use the Adam \cite{DBLP:journals/corr/KingmaB14} optimizer. The initial learning rate is set to 0.001, and the momentum is set to 0.9. The training lasts 20 epochs for all FDPM-related models, and the batch size is chosen as 512. All other parameters are preserved as defaults within the PyTorch \cite{DBLP:conf/nips/PaszkeGMLBCKLGA19} deep learning framework.

\subsection{Results on sampling rates}
\label{sec:res-sr}

\begin{figure}[htb]
	\centering
	\begin{subfigure}{0.38\textwidth}
		\includegraphics[width=\textwidth]{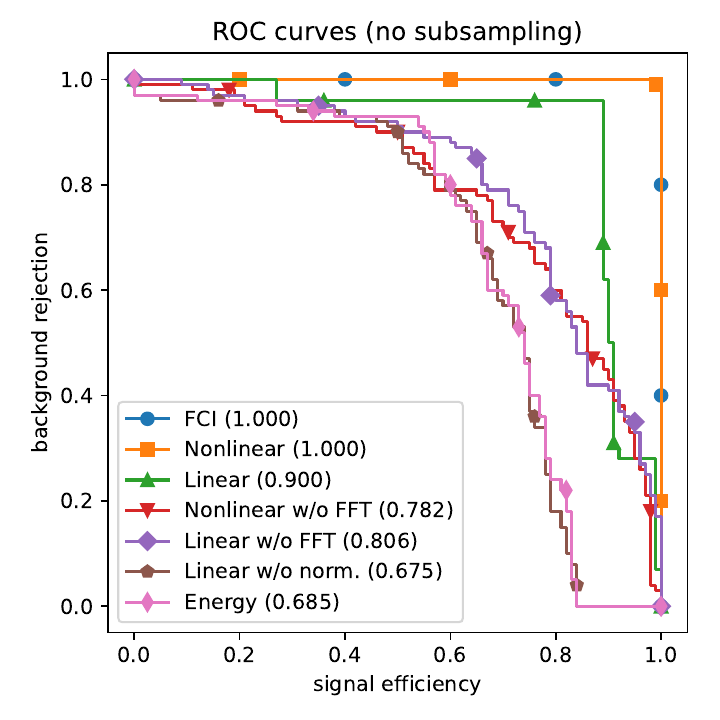}
		\caption{}
		\label{fig:roc-sr-no-sub}
	\end{subfigure}
	\begin{subfigure}{0.38\textwidth}
		\includegraphics[width=\textwidth]{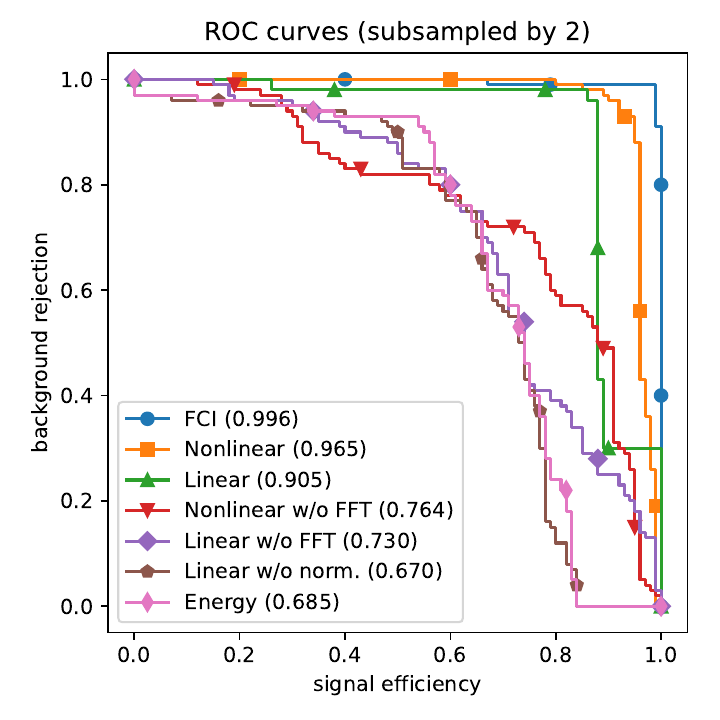}
		\caption{}
		\label{fig:roc-sr-sub2}
	\end{subfigure}
	\begin{subfigure}{0.38\textwidth}
		\includegraphics[width=\textwidth]{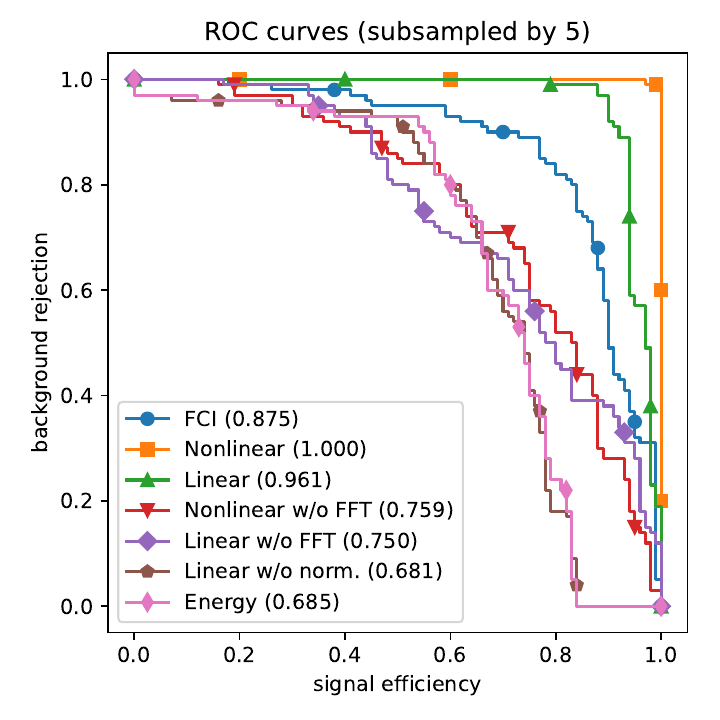}
		\caption{}
		\label{fig:roc-sr-sub5}
	\end{subfigure}
	\begin{subfigure}{0.38\textwidth}
		\includegraphics[width=\textwidth]{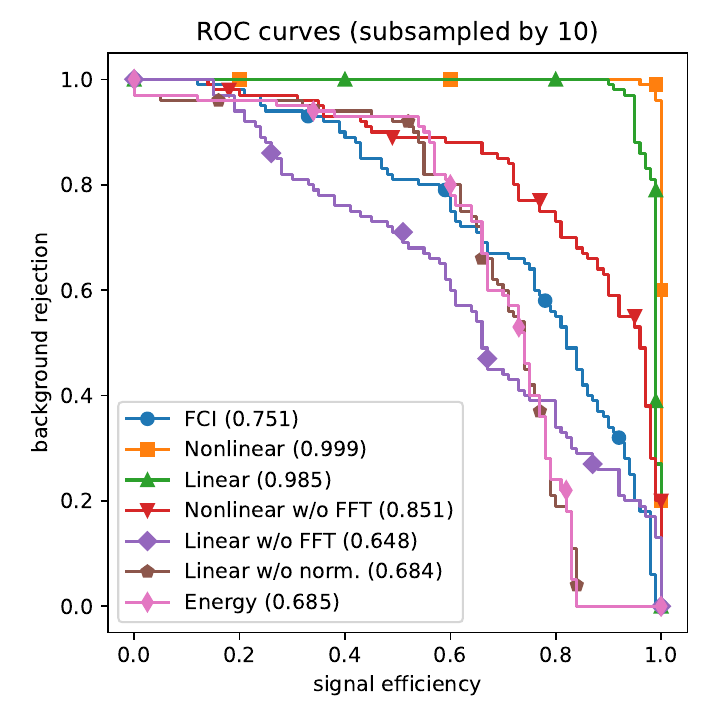}
		\caption{}
		\label{fig:roc-sr-sub10}
	\end{subfigure}
	\caption{ROC curves by different discrimination methods at sampling rates of 100 MS/s, 50 MS/s, 20 MS/s and 10 MS/s. The numbers in the parentheses indicate the area under curve (AUC).}
	\label{fig:roc-sr}
\end{figure}

The original waveform in the dataset is recorded at the sampling rate of 100 MS/s. To study the performance of discriminators at different sampling rates, we subsample the waveform by 2, 5 and 10. This generates the equivalent sampling rates of 50 MS/s, 20 MS/s and 10 MS/s.

In Fig. \ref{fig:roc-sr}, we show the performance of several neutron/gamma discriminators in the form of ROC curves at different sampling rates. These ROC curves are computed on adversarial examples discussed in section \ref{sec:roc-adv}. Among the discrimination methods, ``FCI'' stands for the frequency classification index \cite{MORALES2024745}, which is the originally proposed method used with the dataset. We adjust the cut-off frequencies of FCI for best performance according to the sampling rate. ``Nonlinear'' and ``Linear'' stand for FDPMs proposed in this paper. Apart from the whole forms, we also experiment with models without FFT and models without charge normalization for ablation study, and these models are indicated with ``w/o FFT'' and ``w/o norm''. At last, ``Energy'' stands for a discriminator by charge integration before normalization, which serves as a baseline in experiments. We have calculated the area under curve (AUC) for each discrimination method and annotated it in the parentheses after the corresponding item in the figure legend. The larger AUC indicates better discrimination ability.

It can be seen in Fig. \ref{fig:roc-sr} that different discrimination methods generate varied ROC curves on adversarial examples. FCI is a nearly perfect discriminator at relatively high sampling rates of 100 MS/s and 50 MS/s (Fig. \ref{fig:roc-sr-no-sub} and Fig. \ref{fig:roc-sr-sub2}), but its performance has dropped at relatively low sampling rates of 20 MS/s and 10 MS/s (Fig. \ref{fig:roc-sr-sub5} and Fig. \ref{fig:roc-sr-sub10}). In comparison, the nonlinear FDPM achieves nearly perfect ROC curves at all sampling rates, which is only slightly inferior to FCI at 50 MS/s. For the linear FDPM, its performance is not as good as FCI at relatively high sampling rates, but it is still better than FCI at relatively low sampling rates. In short, nonlinear and linear FDPMs significantly improve discrimination ability and show better adaptability to low sampling rates than FCI.

To assess the necessity of FFT and charge normalization, we also include partial models in Fig. \ref{fig:roc-sr}. It can be seen that, parametric models without FFT, whether it is nonlinear or linear, suffer from significant performance degradation when compared to whole models. This demonstrates FFT is essential for parametric models to achieve superior performance. Besides, we also include the results of the linear model when charge normalization is not performed. Without this preprocessing step, the linear model gives an ROC curve very close to the ROC curve generated by the energy discriminator. This demonstrates that charge normalization is essential for parametric models to learn energy-independent features. Without charge normalization, the energy gap between neutron and gamma events will prevent learning-based models from exploiting pulse characteristics beyond the most prominent feature of energy.

\subsection{Results on noise conditions}
\label{sec:res-noise}

\begin{figure}[htb]
	\centering
	\begin{subfigure}{0.38\textwidth}
		\includegraphics[width=\textwidth]{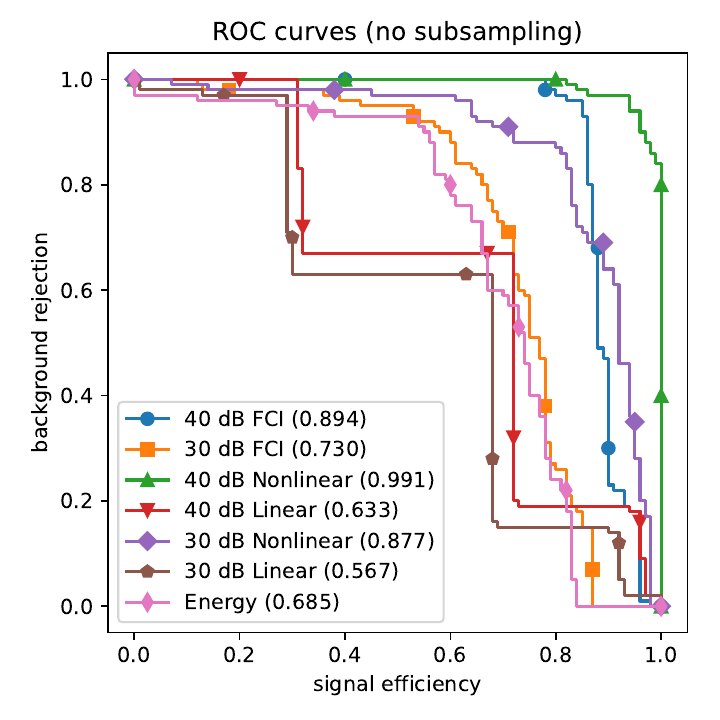}
		\caption{}
		\label{fig:roc-noise-no-sub}
	\end{subfigure}
	\begin{subfigure}{0.38\textwidth}
		\includegraphics[width=\textwidth]{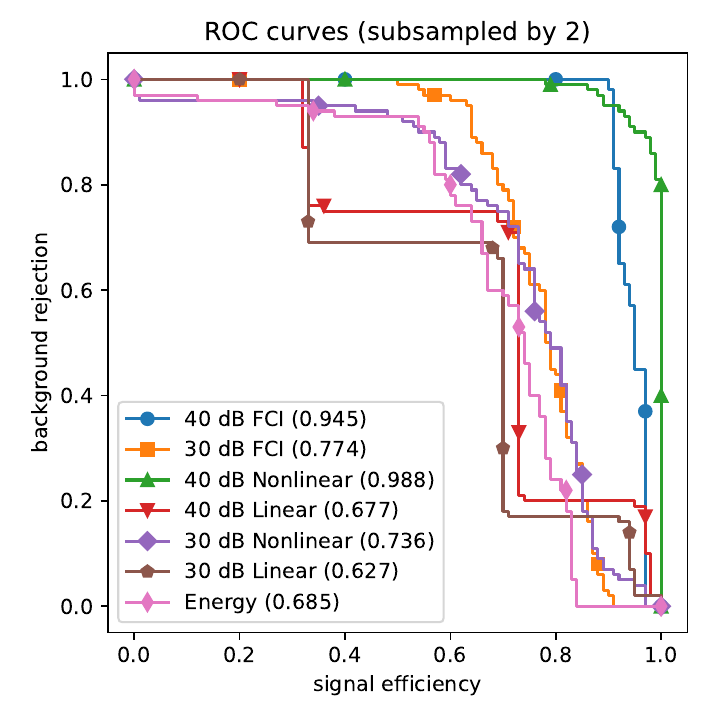}
		\caption{}
		\label{fig:roc-noise-sub2}
	\end{subfigure}
	\begin{subfigure}{0.38\textwidth}
		\includegraphics[width=\textwidth]{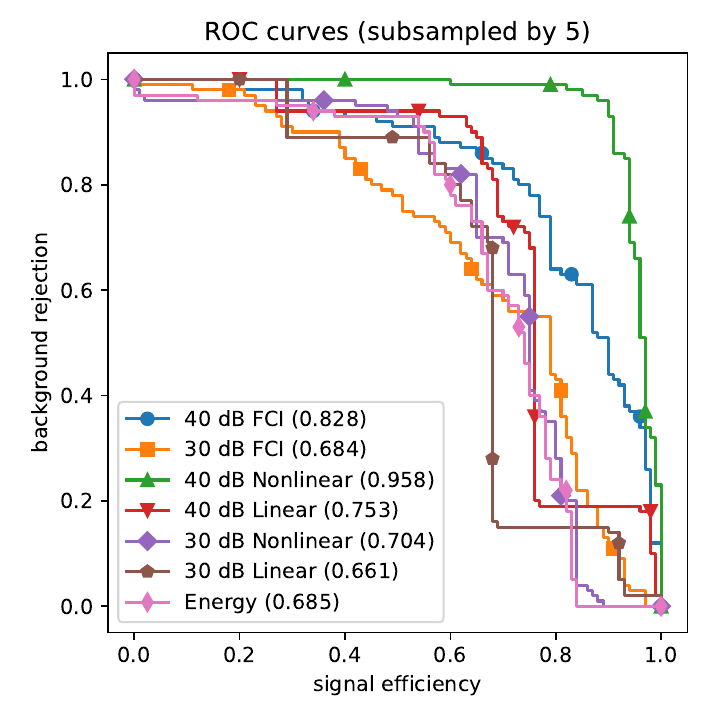}
		\caption{}
		\label{fig:roc-noise-sub5}
	\end{subfigure}
	\begin{subfigure}{0.38\textwidth}
		\includegraphics[width=\textwidth]{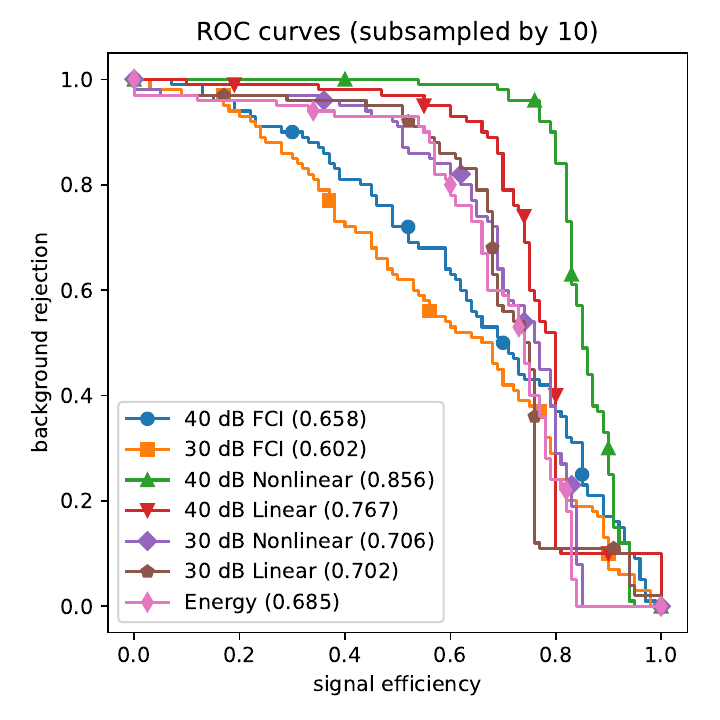}
		\caption{}
		\label{fig:roc-noise-sub10}
	\end{subfigure}
	\caption{ROC curves by different discrimination methods at sampling rates of 100 MS/s, 50 MS/s, 20 MS/s and 10 MS/s, with different noise levels (30 dB/40 dB relative to the maximum amplitude in the dataset). The numbers in the parentheses indicate the area under curve (AUC).}
	\label{fig:roc-noise}
\end{figure}

In this experiment, we apply extra noise on adversarial examples to investigate performance of discriminators under conditions where noise is enhanced. Firstly, we research Gaussian white noise at the side of electronics. Two levels of noise, with SNR of 30 dB and 40 dB relative to the maximum amplitude in the dataset, are added to waveforms. Noise of 30-dB SNR is overwhelmingly high to suppress small pulses, and noise of 40-dB SNR is weaker but still significant.

In Fig. \ref{fig:roc-noise}, we show the ROC curves of neutron/gamma discriminators with two noise levels at different sampling rates. It can be seen that the nonlinear FDPM achieves the best ROC curves across noise levels and sampling rates. When the sampling rate becomes lower, the ROC curve tends to become worse, and the impact of sampling rates on FCI is much heavier than the nonlinear models. For the linear FDPM, it shows a reverse tendency that the AUCs are better with lower sampling rates. Higher sampling rates will generate more data points feeding the models, and simple linear models may poorly converge on information not helpful to distinguish between adversarial examples.

\begin{table}[h!]
	\centering
	\caption{\label{tab:nn-metrics} Evaluation metrics computed on all data in the test dataset and specific data with adversarial sampling. The upper part is related to section \ref{sec:res-sr}, and the lower part is related to section \ref{sec:res-noise}.}
	\footnotesize
	\begin{tabular}{c|c|ccc|ccc}
		\hline
		\multicolumn{2}{c}{} & \multicolumn{3}{c}{all} & \multicolumn{3}{c}{adversarial} \\
		\multicolumn{2}{c}{} & Errors & Acc.(\%) & AUC & Errors & Acc.(\%) & AUC \\
		\hline
		\multirow{7}{*}{no sub.} & FCI & 2 & 99.97 & 1.000 & 1 & 99.50 & 1.000 \\
		& Nonlinear & 3 & 99.96 & 1.000 & 2 & 99.00 & 1.000 \\
		& Linear & 19 & 99.75 & 0.998 & 15 & 92.50 & 0.900 \\
		& Nonlinear w/o FFT & 572 & 92.59 & 0.948 & 65 & 67.50 & 0.782 \\
		& Linear w/o FFT & 1213 & 84.29 & 0.805 & 66 & 67.00 & 0.806 \\
		& Linear w/o norm. & 59 & 99.24 & 0.999 & 59 & 70.50 & 0.675 \\
		& Energy & 53 & 99.31 & 0.999 & 53 & 73.50 & 0.685 \\
		\hline
		\multirow{7}{*}{sub. by 2} & FCI & 52 & 99.33 & 0.998 & 3 & 98.50 & 0.996 \\
		& Nonlinear & 14 & 99.82 & 1.000 & 12 & 94.00 & 0.965 \\
		& Linear & 19 & 99.75 & 0.998 & 16 & 92.00 & 0.905 \\
		& Nonlinear w/o FFT & 549 & 92.89 & 0.944 & 61 & 69.50 & 0.764 \\
		& Linear w/o FFT & 1356 & 82.44 & 0.787 & 83 & 58.50 & 0.730 \\
		& Linear w/o norm. & 59 & 99.24 & 0.999 & 59 & 70.50 & 0.670 \\
		& Energy & 53 & 99.31 & 0.999 & 53 & 73.50 & 0.685 \\
		\hline
		\multirow{7}{*}{sub. by 5} & FCI & 1094 & 85.83 & 0.793 & 51 & 74.50 & 0.875 \\
		& Nonlinear & 4 & 99.95 & 1.000 & 3 & 98.50 & 1.000 \\
		& Linear & 24 & 99.69 & 0.999 & 13 & 93.50 & 0.961 \\
		& Nonlinear w/o FFT & 610 & 92.10 & 0.936 & 70 & 65.00 & 0.759 \\
		& Linear w/o FFT & 1562 & 79.77 & 0.746 & 84 & 58.00 & 0.750 \\
		& Linear w/o norm. & 57 & 99.26 & 0.999 & 57 & 71.50 & 0.681 \\
		& Energy & 53 & 99.31 & 0.999 & 53 & 73.50 & 0.685 \\
		\hline
		\multirow{7}{*}{sub. by 10} & FCI & 1610 & 79.15 & 0.646 & 77 & 61.50 & 0.751 \\
		& Nonlinear & 4 & 99.95 & 1.000 & 3 & 98.50 & 0.999 \\
		& Linear & 16 & 99.79 & 1.000 & 8 & 96.00 & 0.985 \\
		& Nonlinear w/o FFT & 828 & 89.28 & 0.880 & 50 & 75.00 & 0.851 \\
		& Linear w/o FFT & 1681 & 78.23 & 0.709 & 85 & 57.50 & 0.648 \\
		& Linear w/o norm. & 56 & 99.27 & 0.999 & 56 & 72.00 & 0.684 \\
		& Energy & 53 & 99.31 & 0.999 & 53 & 73.50 & 0.685 \\
		\hline
		\hline
		\multirow{7}{*}{no sub.} & 40 dB FCI & 19 & 99.75 & 1.000 & 19 & 90.50 & 0.894 \\
		& 30 dB FCI & 50 & 99.35 & 1.000 & 50 & 75.00 & 0.730 \\
		& 40 dB Nonlinear & 11 & 99.86 & 1.000 & 9 & 95.50 & 0.991 \\
		& 40 dB Linear & 61 & 99.21 & 0.994 & 61 & 69.50 & 0.633 \\
		& 30 dB Nonlinear & 39 & 99.50 & 1.000 & 32 & 84.00 & 0.877 \\
		& 30 dB Linear & 69 & 99.11 & 0.993 & 69 & 65.50 & 0.567 \\
		& Energy & 53 & 99.31 & 0.999 & 53 & 73.50 & 0.685 \\
		\hline
		\multirow{7}{*}{sub. by 2} & 40 dB FCI & 14 & 99.82 & 1.000 & 10 & 95.00 & 0.945 \\
		& 30 dB FCI & 46 & 99.40 & 1.000 & 46 & 77.00 & 0.774 \\
		& 40 dB Nonlinear & 20 & 99.74 & 1.000 & 14 & 93.00 & 0.988 \\
		& 40 dB Linear & 57 & 99.26 & 0.995 & 56 & 72.00 & 0.677 \\
		& 30 dB Nonlinear & 53 & 99.31 & 0.999 & 53 & 73.50 & 0.736 \\
		& 30 dB Linear & 63 & 99.18 & 0.994 & 63 & 68.50 & 0.627 \\
		& Energy & 53 & 99.31 & 0.999 & 53 & 73.50 & 0.685 \\
		\hline
		\multirow{7}{*}{sub. by 5} & 40 dB FCI & 403 & 94.78 & 0.953 & 59 & 70.50 & 0.828 \\
		& 30 dB FCI & 125 & 98.38 & 0.992 & 79 & 60.50 & 0.684 \\
		& 40 dB Nonlinear & 17 & 99.78 & 1.000 & 15 & 92.50 & 0.958 \\
		& 40 dB Linear & 44 & 99.43 & 0.997 & 44 & 78.00 & 0.753 \\
		& 30 dB Nonlinear & 53 & 99.31 & 0.999 & 53 & 73.50 & 0.704 \\
		& 30 dB Linear & 55 & 99.29 & 0.996 & 55 & 72.50 & 0.661 \\
		& Energy & 53 & 99.31 & 0.999 & 53 & 73.50 & 0.685 \\
		\hline
		\multirow{7}{*}{sub. by 10} & 40 dB FCI & 1180 & 84.72 & 0.757 & 93 & 53.50 & 0.658 \\
		& 30 dB FCI & 501 & 93.51 & 0.922 & 95 & 52.50 & 0.602 \\
		& 40 dB Nonlinear & 27 & 99.65 & 1.000 & 27 & 86.50 & 0.856 \\
		& 40 dB Linear & 43 & 99.44 & 0.998 & 42 & 79.00 & 0.767 \\
		& 30 dB Nonlinear & 56 & 99.27 & 0.999 & 56 & 72.00 & 0.706 \\
		& 30 dB Linear & 52 & 99.33 & 0.997 & 52 & 74.00 & 0.702 \\
		& Energy & 53 & 99.31 & 0.999 & 53 & 73.50 & 0.685 \\
		\hline
	\end{tabular}
\end{table}

In table \ref{tab:nn-metrics}, we list the number of misclassified examples (errors), classification accuracy and AUC for all data and specific data with adversarial sampling, with respect to all cases considered in section \ref{sec:res-sr} and section \ref{sec:res-noise}. The following points are made according to the results:

\begin{itemize}
	\item Apart from models without FFT and FCI at relatively low sampling rates, almost all discriminators achieve nearly perfect metrics on all data in the test dataset. However, on specific data with adversarial sampling, different methods have exhibited noticeable diversity, which can be used for precision assessments on discriminators.
	\item For most nearly perfect discriminators (errors below 100 on all data), the errors on adversarial examples take a large portion of all errors, which demonstrates the effectiveness of adversarial sampling.
	\item For discriminators not good enough, errors on adversarial examples may be underestimated when compared to errors on all data. This demonstrates adversarial sampling can be more justified within the scope of high-precision discriminators.
\end{itemize}

Besides, FDPM gives the best overall performance, and errors of these models on adversarial examples show good coincidence with errors on all data. Therefore, these models can be well judged by the proposed evaluation criterion.

\subsection{Additional results on noise complexity}
\label{sec:add-noise-com}

\begin{figure}[htb]
	\centering
	\begin{subfigure}{0.38\textwidth}
		\includegraphics[width=\textwidth]{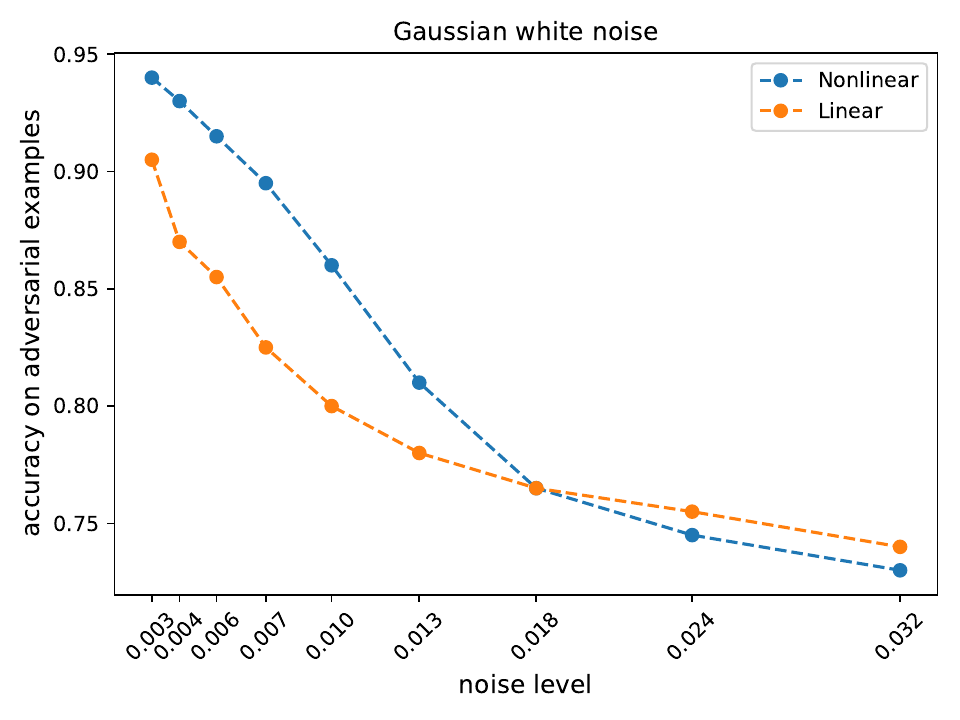}
		\caption{}
		\label{fig:gaus-white-scan}
	\end{subfigure}
	\begin{subfigure}{0.38\textwidth}
		\includegraphics[width=\textwidth]{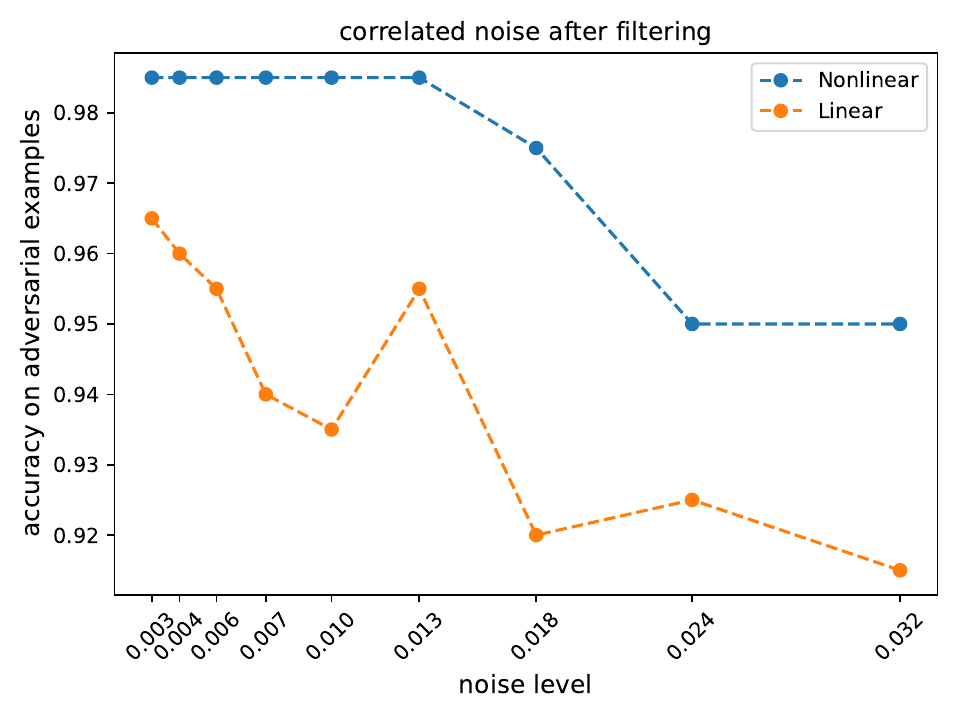}
		\caption{}
		\label{fig:corr-scan}
	\end{subfigure}
	\begin{subfigure}{0.38\textwidth}
		\includegraphics[width=\textwidth]{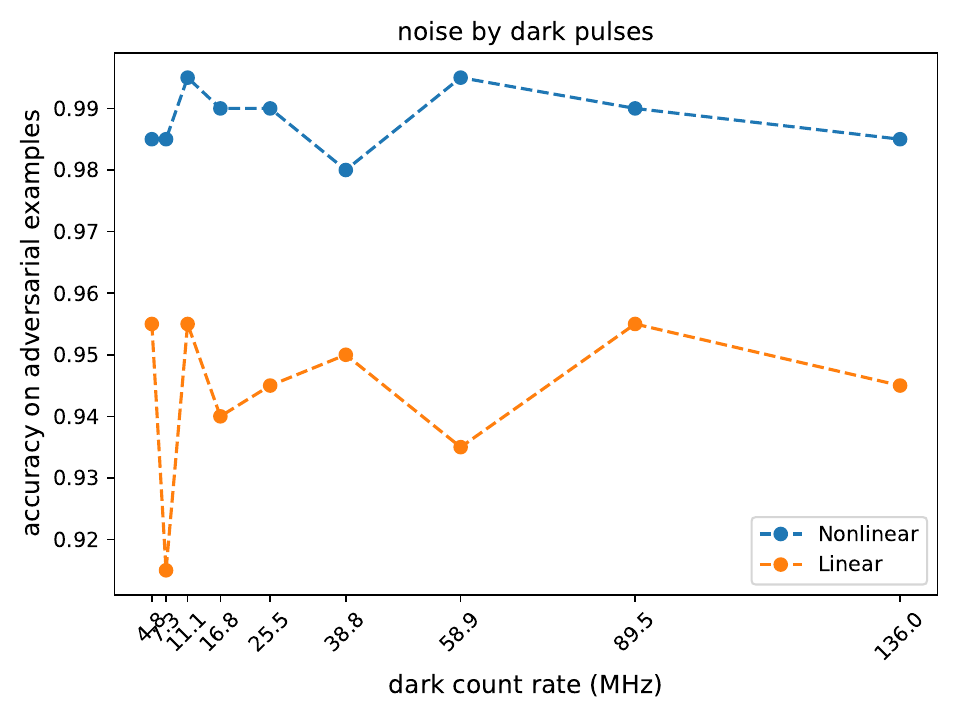}
		\caption{}
		\label{fig:dark-pulse-scan}
	\end{subfigure}
	\caption{The performance sensitivity of the proposed method on different noise components. (a) Gaussian white noise. (b) Correlated noise after band-limited filtering. (c) Noise incurred by dark pulses.}
	\label{fig:noise-scan}
\end{figure}

We continue to discuss the effects of noise in some more complex forms. According to the specification of the detector \cite{detector-frontend}, the rise time $t_r$ of pulses induced by gammas is 1.5 $\mu$s. Considering the rise time on SiPMs is much shorter than that, the bandwidth of the detector is estimated by the empirical equation to be $0.35/t_r = 233.3\ \mathrm{kHz}$. Hence, we use this bandwidth to generate correlated noise by Gaussian random process. The accuracy on adversarial examples vs. the noise level (standard deviation of noise divided by the maximum signal amplitude) is shown in Fig. \ref{fig:corr-scan}. When choosing the noise level, we linearly sample values on the logarithm scale from 50 dB to 30 dB. It can be seen that the performance of linear and nonlinear FDPMs tend to degrade at higher noise level. However, when comparing to the Gaussian white noise in Fig. \ref{fig:gaus-white-scan} at the same noise level, the performance degradation is much less.

Regarding the noise incurred by dark pulses, we refer to \cite{CLYC-scint} for the light yield of about 20000 photons/MeV of gamma energy by the CLYC crystal. According to the linear relation $E = 9.024 \cdot a\ \mathrm{(keVee)}$ between the gamma energy $E$ and the ADC count $a$ \cite{MORALES2024745}, the peak amplitude of dark pulses (single p.e.) is estimated to be $1 / 20000 * 1000 / 9.024 = 0.0055\ \mathrm{(ADC\ counts)}$. Besides, we refer to \cite{MICROC-SiPM} for the dark count rate of the MicroC-60035 SiPM to be 1.2 MHz (typ.)/3.4 MHz (max.), and the array of 4 SiPMs should be four times as high. With the above information, the effective values of dark pulses are generated by convolving pulse counts at each time step (from a Poisson distribution) and the single p.e. waveform.

In Fig. \ref{fig:dark-pulse-scan}, we plot the accuracy on adversarial examples at different dark count rates. The dark count rates are chosen from the typical value to ten times the maximum value, linearly on a logarithm scale. Considering the limited bandwidth, the actual effect of dark pulses is to raise the baseline with minor fluctuations. Both linear and nonlinear FDPMs show slightly varied performance at low dark count rates, but their accuracies do not degrade on a larger scale. In general, the proposed method exhibits robustness to correlated noise and dark pulses with the CLYC detector in use.

\subsection{Comparison to other methods}

\begin{table}[h!]
	\centering
	\caption{\label{tab:cnn-lstm-arch} Network architectures of CNN and LSTM models. ``k'' stands for kernel size, ``s'' stands for stride, and ``d'' stands for dimension of the feature map. For CNN and LSTM, the ReLU activation follows each convolution or linear layer, except for the last linear layer.}
	\footnotesize
	\begin{tabular}{c|c}
		\hline
		Model Name & Architecture \\
		\hline
		CNN & \makecell{Conv(k:6, s:1) $\rightarrow$ MaxPool(k:4) $\rightarrow$ Conv(k:4, s:1) $\rightarrow$ MaxPool(k:3) $\rightarrow$ Conv(k:4, s:1) \\ $\rightarrow$ MaxPool(k:2) $\rightarrow$ Flatten $\rightarrow$ Linear(d:64) $\rightarrow$ Linear(d:2)} \\
		\hline
		LSTM & LSTMCell(k:64) $\rightarrow$ LastOutput $\rightarrow$ Linear(d:32) $\rightarrow$ Linear(d:2) \\
		\hline
	\end{tabular}
\end{table}

\begin{table}[h!]
	\centering
	\caption{\label{tab:algo-metrics} Evaluation metrics computed on all data in the test dataset and specific data with adversarial sampling, at the sampling rate of 10 MS/s, by novel machine learning and conventional discrimination methods. Upper part: with charge normalization. Lower part: with min-max normalization.}
	\footnotesize
	\begin{tabular}{c|c|ccc|ccc}
		\hline
		\multicolumn{2}{c}{} & \multicolumn{3}{c}{all} & \multicolumn{3}{c}{adversarial} \\
		\multicolumn{2}{c}{} & Errors & Acc.(\%) & AUC & Errors & Acc.(\%) & AUC \\
		\hline
		\multirow{2}{*}{w/ charge norm.} & CNN & 36 & 99.53 & 1.000 & 32 & 84.00 & 0.906 \\
        & LSTM & 886 & 88.53 & 0.859 & 91 & 54.50 & 0.498 \\
		\hline
		\multirow{6}{*}{w/ min-max norm.} & Charge Comparison & 710 & 90.81 & 0.915 & 87 & 56.50 & 0.685 \\
        & Zero Crossing & 846 & 89.05 & 0.888 & 98 & 51.00 & 0.540 \\
        & PCNN & 2167 & 71.94 & 0.157 & 101 & 49.50 & 0.346 \\
        & Ladder Gradient & 2009 & 73.99 & 0.323 & 95 & 52.50 & 0.550 \\
        & FGA & 1661 & 78.49 & 0.699 & 101 & 49.50 & 0.446 \\
        & FEPS & 1193 & 84.55 & 0.742 & 82 & 59.00 & 0.699 \\
		\hline
	\end{tabular}
\end{table}

\begin{figure}[htb]
	\centering
	\includegraphics[width=0.38\textwidth]{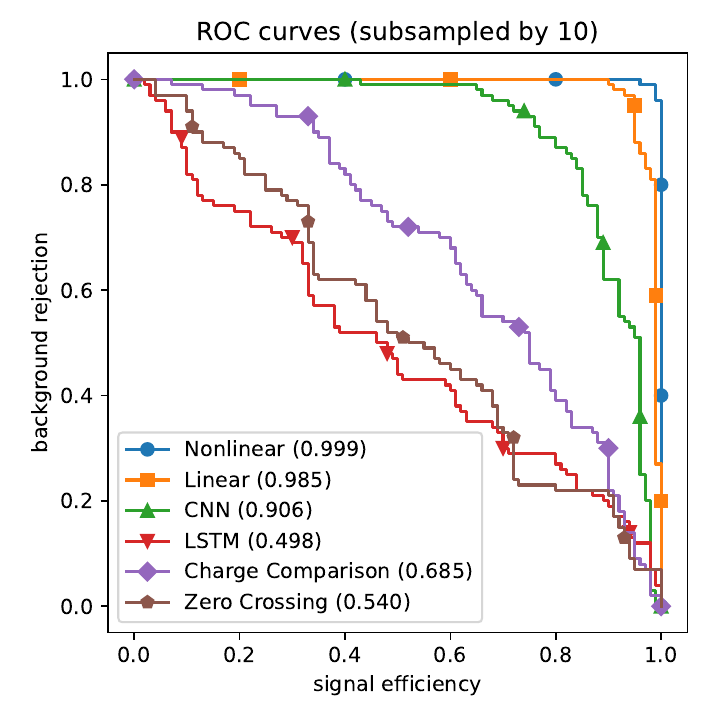}
	\caption{ROC curves by different discrimination methods at the sampling rate of 10 MS/s. The numbers in the parentheses indicate the area under curve (AUC).}
	\label{fig:roc-algo}
\end{figure}

We test two novel ML methods, the convolutional neural network (CNN) \cite{YOON20233925} and the long short-term memory (LSTM) \cite{FABIAN2021164750}, both of which have advanced network architectures. The architectures of these two models are listed in table \ref{tab:cnn-lstm-arch}. Besides, we adjust neutron/gamma discriminators implemented in \cite{NGMethods} and apply them to the dataset used above. Six conventional methods are tested in total: charge comparison, zero crossing, pulse-coupled neural network (PCNN), ladder gradient, frequency gradient analysis (FGA) and falling-edge percentage slope (FEPS).

For CNN and LSTM, we test them with charge normalization just as FDPM. For conventional methods, we test them with min-max normalization to restrict the amplitudes in the range (0, 1). In table \ref{tab:algo-metrics}, it can be seen that CNN gives competitive performance against FDPM; on adversarial examples, however, its accuracy and AUC are worse than FDPM. Among other methods, LSTM, charge comparison and zero crossing give fair performance well above random guesses, but their accuracies are not as good as FDPM and CNN. We select these discussed methods, along with linear and nonlinear FDPMs, to plot their ROC curves, shown in Fig. \ref{fig:roc-algo}. Different methods exhibit diverse qualities in these curves, from nearly neutral performance (straight line from (0, 1) to (1, 0)) to nearly perfect performance (broken line from (0, 1) to (1, 1) and to (1, 0)). Under the setting of the CLYC dataset, FDPM is a preferable choice as a top-ranking neutron/gamma discriminator.

\subsection{Additional results on a TOF-labelled organic dataset}
\label{sec:add-tof}

To more comprehensively and objectively evaluate different discrimination methods, and to give useful clues about the adversarial sampling strategy, we also include a brief discussion on a neutron/gamma dataset with the organic scintillator \cite{TOFDataset}. In this dataset, neutron events and gamma events are labelled by time-of-flight (TOF) differences of these particles. The sampling rate is much higher (500 MS/s), and the signal pulse is much sharper, reflecting the high bandwidth of the detector. The data examples are subsampled with a random starting point from 0 to 9, and the first 128 data points, capable of covering all valid pulses, are selected with no downsampling. Normalization is performed according to the used discrimination method. When training ML models, we use 5-fold cross validation since the examples are much fewer (2197 neutron events and 2197 gamma events). The test sets in each fold are then combined together to form a whole test dataset.

\begin{figure}[htb]
	\centering
	\begin{subfigure}{0.38\textwidth}
		\includegraphics[width=\textwidth]{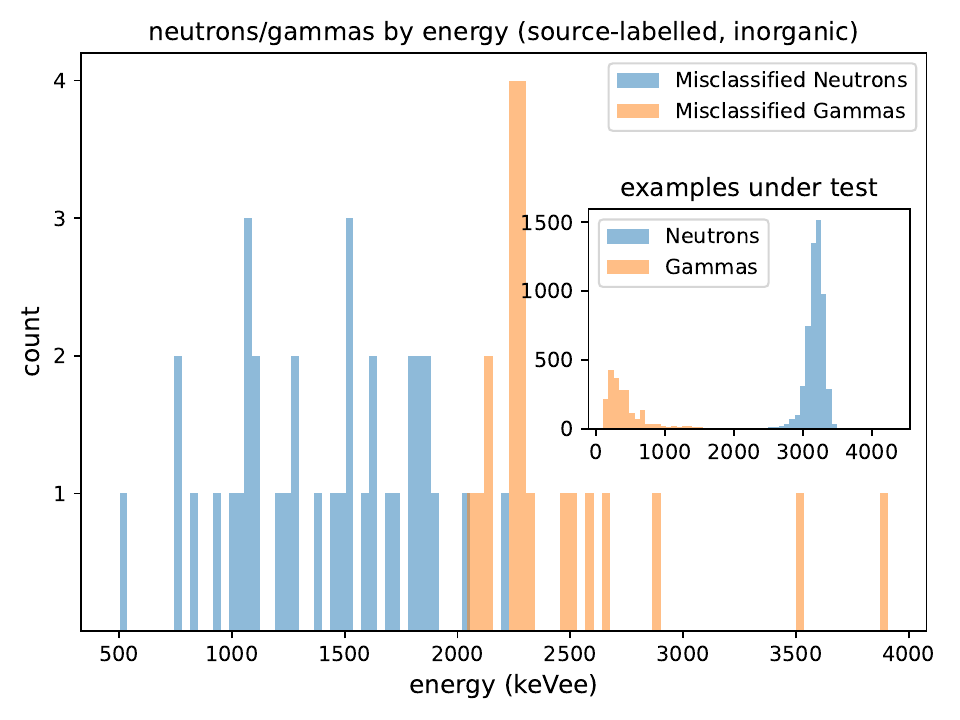}
		\caption{}
		\label{fig:source-ng-dist}
	\end{subfigure}
	\begin{subfigure}{0.38\textwidth}
		\includegraphics[width=\textwidth]{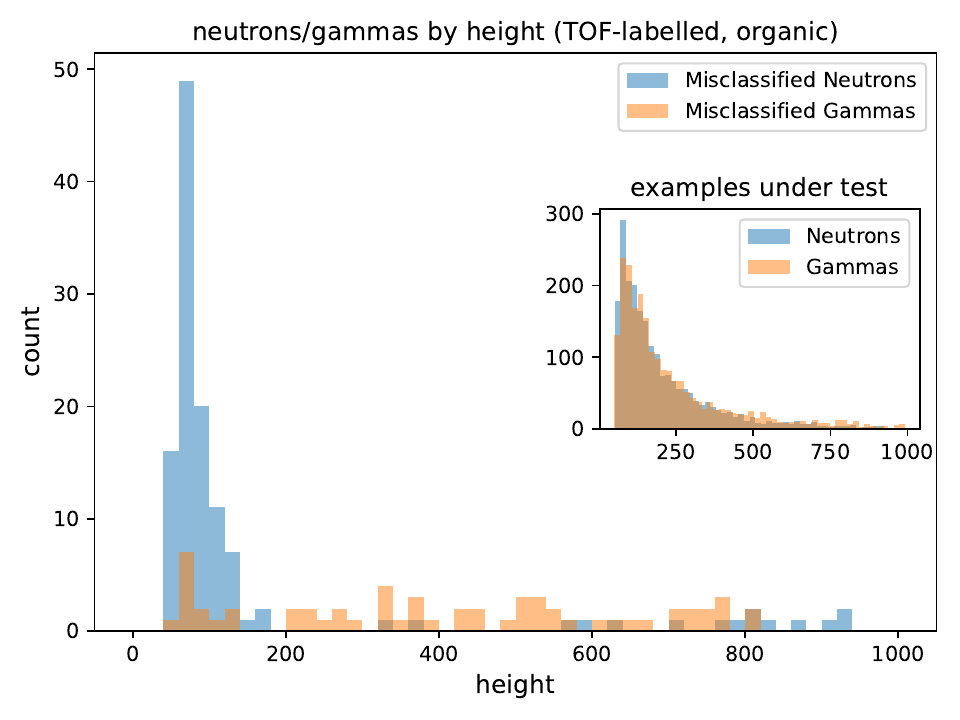}
		\caption{}
		\label{fig:tof-ng-error-dist}
	\end{subfigure}
	\caption{Histograms illustrating the distributions of misclassified examples over energy/height. The distributions of all test examples are also plotted as inset axes. (a) Source-labelled CLYC dataset. (b) TOF-labelled organic dataset.}
	\label{fig:ng-error-dist}
\end{figure}

\begin{table}[h!]
	\centering
	\caption{\label{tab:add-algo-metrics} Additional evaluation metrics computed on all data in the test dataset and specific data with adversarial sampling, for the organic dataset, by different discrimination methods. Upper part: with charge normalization. Lower part: with min-max normalization.}
	\footnotesize
	\begin{tabular}{c|c|ccc|ccc}
		\hline
		\multicolumn{2}{c}{} & \multicolumn{3}{c}{all} & \multicolumn{3}{c}{adversarial} \\
		\multicolumn{2}{c}{} & Errors & Acc.(\%) & AUC & Errors & Acc.(\%) & AUC \\
		\hline
		\multirow{5}{*}{w/ charge norm.} & FCI & 1209 & 72.49 & 0.776 & 331 & 66.90 & 0.707 \\
        & Nonlinear & 183 & 95.84 & 0.984 & 170 & 83.00 & 0.872 \\
        & Linear & 180 & 95.90 & 0.986 & 168 & 83.20 & 0.874 \\
        & CNN & 184 & 95.81 & 0.982 & 169 & 83.10 & 0.860 \\
        & LSTM & 997 & 77.31 & 0.791 & 327 & 67.30 & 0.712 \\
		\hline
		\multirow{6}{*}{w/ min-max norm.} & Charge Comparison & 179 & 95.93 & 0.993 & 168 & 83.20 & 0.916 \\
        & Zero Crossing & 403 & 90.83 & 0.967 & 289 & 71.10 & 0.819 \\
        & PCNN & 254 & 94.22 & 0.966 & 217 & 78.30 & 0.807 \\
        & Ladder Gradient & 1568 & 64.31 & 0.438 & 322 & 67.80 & 0.612 \\
        & FGA & 1408 & 67.96 & 0.742 & 425 & 57.50 & 0.634 \\
        & FEPS & 735 & 83.27 & 0.892 & 249 & 75.10 & 0.789 \\
		\hline
	\end{tabular}
\end{table}

\begin{figure}[htb]
	\centering
	\includegraphics[width=0.38\textwidth]{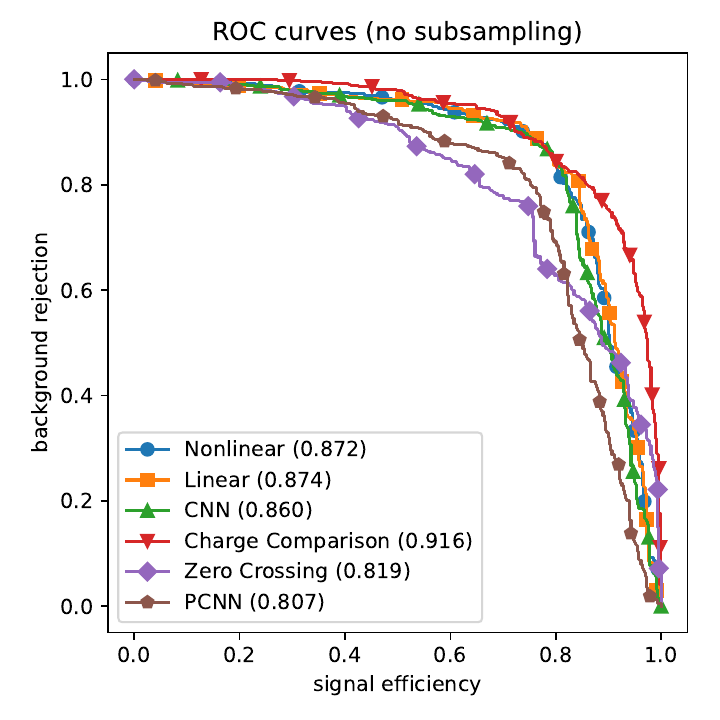}
	\caption{Additional ROC curves by different discrimination methods for the organic dataset. The numbers in the parentheses indicate the area under curve (AUC).}
	\label{fig:add-roc-algo}
\end{figure}

Since the characteristics of organic dataset are very different from the former CLYC dataset, the critical examples might have a different energy distribution. In Fig. \ref{fig:ng-error-dist}, we plot the energy distribution of misclassified examples by FDPM in these two datasets. It can be seen that misclassified examples in the CLYC dataset have apparent energy dependencies; however, for the organic dataset, misclassified neutrons are concentrated in the low energy region, while misclassified gammas spread in the full energy range. As a result, the adversarial sampling strategy must be adjusted for the organic dataset.

Since charge comparison is a simple and interpretable discrimination method, we use the PSD index provide by charge comparison and select 1000 examples near the PSD crossing point as our adversarial examples. The evaluation results are shown in table \ref{tab:add-algo-metrics}. LSTM faces the convergence issue that only 2 out of 5 folds have successfully converged after training lasts for sufficient epochs, so its results are partially valid. It can be seen that nonlinear and linear FDPMs, CNN and charge comparison achieve best accuracies and AUCs. On adversarial examples covering most of their errors, charge comparison is slightly better in the AUC metric, but the accuracy metric is very close for these methods. Among other methods, zero crossing and PCNN show fair performance with lower accuracies. We plot the ROC curves of above discussed methods, shown in Fig. \ref{fig:add-roc-algo}. The differences between these methods are not as large as the CLYC dataset, and optimization-based methods (FDPM and CNN) give very similar ROC curves. Validated on the organic dataset, FDPM adapts to different data distributions and shows justifiable discrimination power.

\subsection{Model interpretation}

\begin{figure}[htb]
	\centering
	\begin{subfigure}{0.38\textwidth}
		\includegraphics[width=\textwidth]{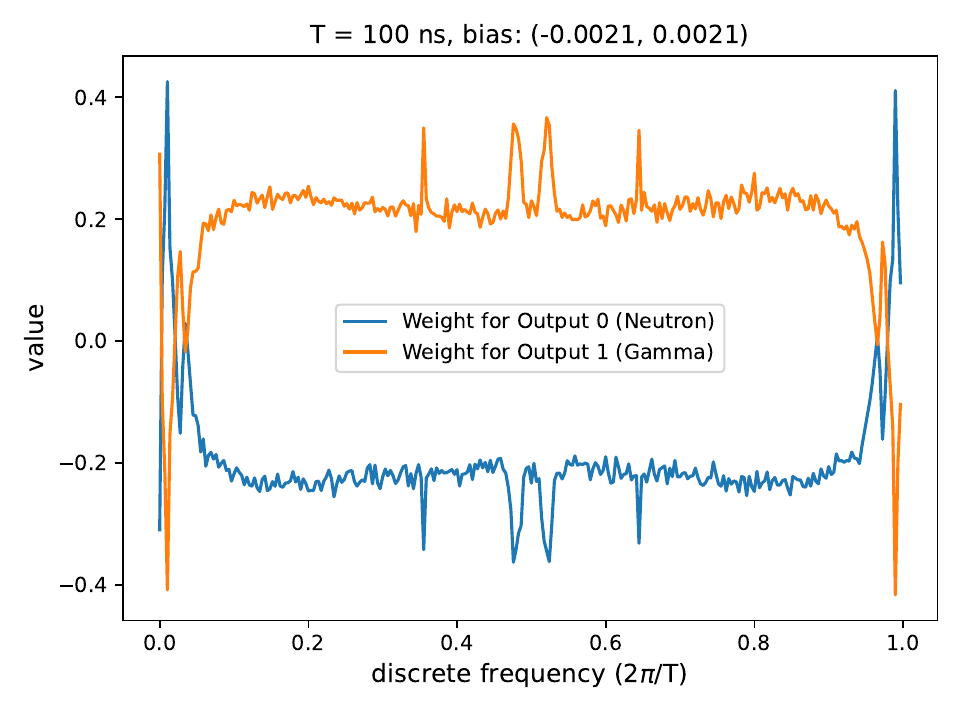}
		\caption{}
		\label{fig:source-weight-visual}
	\end{subfigure}
	\begin{subfigure}{0.38\textwidth}
		\includegraphics[width=\textwidth]{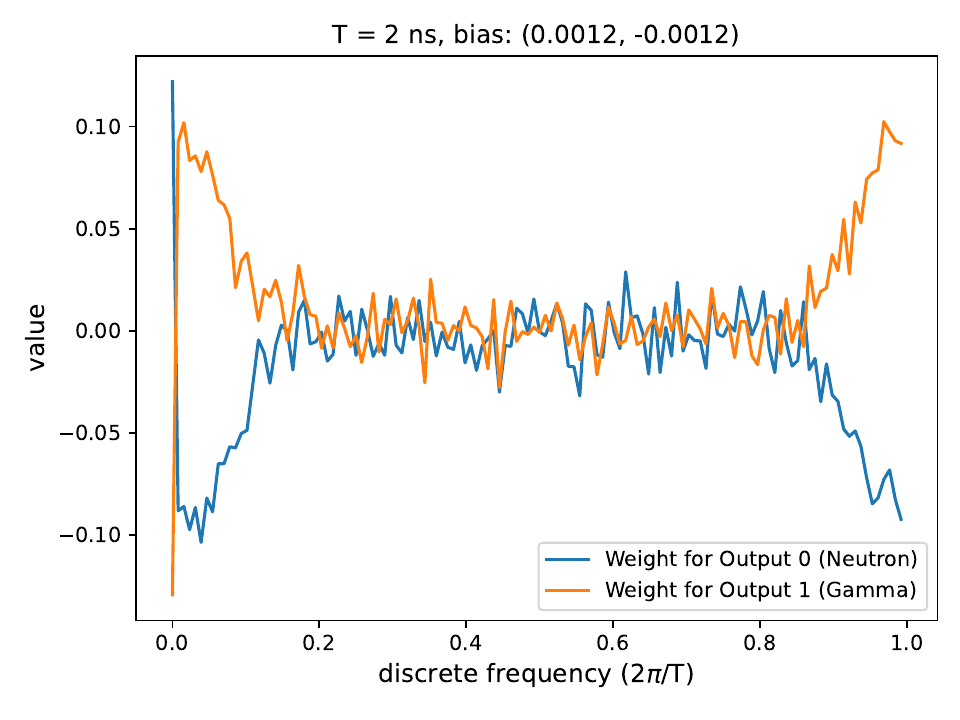}
		\caption{}
		\label{fig:tof-weight-visual}
	\end{subfigure}
	\caption{Visualizing weights of the linear FDPM on the discrete spectrum. (a) CLYC dataset with the 10 MS/s sampling rate. (b) Organic dataset with the 500 MS/s sampling rate.}
	\label{fig:weight-visual}
\end{figure}

A noteworthy advantage of the linear FDPM is its interpretability. In Fig. \ref{fig:weight-visual}, we separate the weight matrices of two targets (neutron and gamma) and plot the weight vector for each target on the discrete spectrum. It can be seen that the exceptional weights lie heavily in the low frequency region (and the mirroring high frequency region since the spectrum is periodic). There is a turning point at the zero-frequency, because we do not perform specialized baseline cancellation so that zero-frequency has weakened weights. Since pulses are much sharper in the organic dataset, the distribution of exceptional weights is wider in Fig. \ref{fig:tof-weight-visual} than in Fig. \ref{fig:source-weight-visual}. We also notice some peaks near $\pi/T$-frequency in Fig. \ref{fig:source-weight-visual}, possibly due to slightly over-fitting to noise characteristics in the CLYC dataset. Generally, the weight distribution has similar features as other frequency-domain discrimination methods, but the weight values for each frequency point have been optimized for best performance.

\section{Discussion}

Based on the above demonstration, the advantages of the proposed evaluation criterion and the learning-based model are sufficiently explored. However, they also face several challenges.

The most notable challenge is the acquisition of the labelled dataset. It is well-known that neutron and gamma events are always accompanied by each other. Hence, normal data acquisition procedure will collect both neutron and gamma events, which makes the subsequent labelling a tough task. Although we can produce separate neutron and gamma datasets through simulation, the gap between simulation and reality becomes another challenge. As a result, it is preferable to collected the labelled dataset as an independent procedure using pure sources or time-of-flight before normal data collection.

Another challenge is about the proper use of adversarial sampling. In this study, we use energy as the criterion for adversarial sampling on the CLYC dataset, and the PSD index of charge comparison on the organic dataset. The optimal choice depends on the characteristics of pulses, as well as energy distribution of the dataset. Besides, the amount of examples to sample is uncertain, which may need to be found by trial and error.

It is worthwhile to note that the proposed model is promising to be implemented on economical hardware and portable devices. For example, at the sampling rate of 10 MS/s, the length of the time series can be as short as 256 points. In Fig. \ref{fig:roc-sr-sub10}, it can be seen that the linear FDPM can still achieve good discrimination. Both FFT and addition/multiplication are affordable at this sampling rate, and can be possibly pipelined by custom logic on field programmable gate arrays.

\section{Conclusion}

In this paper, a simple and effective approach based on FDPM is proposed for high-precision neutron/gamma discrimination. In view of the limitation of FoM, we propose to use ROC curves with adversarial sampling to evaluate the ability of different discriminators. The proposed models achieve significantly better results by evaluating on all data in the test dataset, and especially on specific data with adversarial sampling, than other novel machine learning and conventional methods. The simple and interpretable structure of the model also sheds light on efficient hardware implementation with microcontrollers or logic devices. We hope the proposed evaluation criterion and the discrimination method will benefit studies in this area and provide a guidance for the future direction of performance optimization.

\section*{CRediT authorship contribution statement}

\textbf{Pengcheng Ai:} Conceptualization, Software, Writing - Original Draft, Funding acquisition. \textbf{Hongtao Qin:} Formal analysis, Investigation. \textbf{Xiangming Sun:} Writing - Review \& Editing, Supervision. \textbf{Kaiwen Shang:} Data Curation.

\section*{Declaration of competing interest}

The authors declare that they have no known competing financial interests or personal relationships that could have appeared to influence the work reported in this paper.

\section*{Acknowledgements}

This research is supported in part by the National Natural Science Foundation of China (under Grant No. 12405228), in part by the National Key Research and Development Program of China (under Grant No. 2024YFF0726201), in part by the Fundamental Research Funds for the Central Universities (under Grant No. CCNU23XJ013) and in part by the China Postdoctoral Science Foundation (under Grant No. 2023M731244).



\bibliographystyle{elsarticle-num} 
\bibliography{mybibfile}

\begin{thebibliography}{10}
\expandafter\ifx\csname url\endcsname\relax
  \def\url#1{\texttt{#1}}\fi
\expandafter\ifx\csname urlprefix\endcsname\relax\def\urlprefix{URL }\fi
\expandafter\ifx\csname href\endcsname\relax
  \def\href#1#2{#2} \def\path#1{#1}\fi

\bibitem{Ahnouz01122024}
H.~A. Imane~Ahnouz, R.~Sebihi, {A Review of Neutron–Gamma-Ray Discrimination
  Methods Using Organic Scintillators}, Nuclear Science and Engineering
  198~(12) (2024) 2241--2273.
\newblock \href {https://doi.org/10.1080/00295639.2024.2316946}
  {\path{doi:10.1080/00295639.2024.2316946}}.

\bibitem{SODERSTROM2019238}
P.-A. S{\"o}derström, G.~Jaworski, J.~{Valiente Dobón}, J.~Nyberg,
  J.~Agramunt, G.~{de Angelis}, S.~Carturan, J.~Egea, M.~Erduran,
  S.~Ert{\"u}rk, G.~{de France}, A.~Gadea, A.~Goasduff, V.~González,
  K.~Hady{\'n}ska-Kl{\c e}k, T.~Hüyük, V.~Modamio, M.~Moszynski, A.~{Di
  Nitto}, M.~Palacz, N.~Pietralla, E.~Sanchis, D.~Testov, A.~Triossi,
  R.~Wadsworth, {Neutron detection and {$\gamma$}-ray suppression using
  artificial neural networks with the liquid scintillators BC-501A and BC-537},
  Nuclear Instruments and Methods in Physics Research Section A: Accelerators,
  Spectrometers, Detectors and Associated Equipment 916 (2019) 238--245.
\newblock \href {https://doi.org/10.1016/j.nima.2018.11.122}
  {\path{doi:10.1016/j.nima.2018.11.122}}.

\bibitem{LEE2024169638}
S.~Lee, K.~Ko, G.~Song, W.~Kim, S.~Yoon, C.~Lee, K.~T. Lim, G.~Cho,
  {Investigation of neutron/gamma-ray distribution in SiPM-based pulse shape
  discrimination using EJ-276 plastic scintillators}, Nuclear Instruments and
  Methods in Physics Research Section A: Accelerators, Spectrometers, Detectors
  and Associated Equipment 1066 (2024) 169638.
\newblock \href {https://doi.org/10.1016/j.nima.2024.169638}
  {\path{doi:10.1016/j.nima.2024.169638}}.

\bibitem{WOLSKI1995584}
D.~Wolski, M.~Moszy{\'n}ski, T.~Ludziejewski, A.~Johnson, W.~Klamra,
  {\"O}.~Skeppstedt, Comparison of n-{$\gamma$} discrimination by zero-crossing
  and digital charge comparison methods, Nuclear Instruments and Methods in
  Physics Research Section A: Accelerators, Spectrometers, Detectors and
  Associated Equipment 360~(3) (1995) 584--592.
\newblock \href {https://doi.org/10.1016/0168-9002(95)00037-2}
  {\path{doi:10.1016/0168-9002(95)00037-2}}.

\bibitem{BAYAT2012217}
E.~Bayat, N.~Divani-Vais, M.~Firoozabadi, N.~Ghal-Eh, {A comparative study on
  neutron-gamma discrimination with NE213 and UGLLT scintillators using
  zero-crossing method}, Radiation Physics and Chemistry 81~(3) (2012)
  217--220.
\newblock \href {https://doi.org/10.1016/j.radphyschem.2011.10.016}
  {\path{doi:10.1016/j.radphyschem.2011.10.016}}.

\bibitem{5485131}
G.~Liu, M.~J. Joyce, X.~Ma, M.~D. Aspinall, A digital method for the
  discrimination of neutrons and $\gamma$ rays with organic scintillation
  detectors using frequency gradient analysis, IEEE Transactions on Nuclear
  Science 57~(3) (2010) 1682--1691.
\newblock \href {https://doi.org/10.1109/TNS.2010.2044246}
  {\path{doi:10.1109/TNS.2010.2044246}}.

\bibitem{Liu_2016}
M.-Z. Liu, B.-Q. Liu, Z.~Zuo, L.~Wang, G.-B. Zan, X.-G. Tuo, Toward a fractal
  spectrum approach for neutron and gamma pulse shape discrimination, Chinese
  Physics C 40~(6) (2016) 066201.
\newblock \href {https://doi.org/10.1088/1674-1137/40/6/066201}
  {\path{doi:10.1088/1674-1137/40/6/066201}}.

\bibitem{YE2022166256}
H.~Ye, L.~Chen, X.~Xu, G.~Jin, {Fast FPGA algorithm for neutron–gamma
  discrimination}, Nuclear Instruments and Methods in Physics Research Section
  A: Accelerators, Spectrometers, Detectors and Associated Equipment 1027
  (2022) 166256.
\newblock \href {https://doi.org/10.1016/j.nima.2021.166256}
  {\path{doi:10.1016/j.nima.2021.166256}}.

\bibitem{Liu2021}
H.-R. Liu, Y.-X. Cheng, Z.~Zuo, T.-T. Sun, K.-M. Wang, Discrimination of
  neutrons and gamma rays in plastic scintillator based on pulse-coupled neural
  network, Nuclear Science and Techniques 32~(8) (2021) 82.
\newblock \href {https://doi.org/10.1007/s41365-021-00915-w}
  {\path{doi:10.1007/s41365-021-00915-w}}.

\bibitem{Liu2022}
H.-R. Liu, M.-Z. Liu, Y.-L. Xiao, P.~Li, Z.~Zuo, Y.-H. Zhan, Discrimination of
  neutron and gamma ray using the ladder gradient method and analysis of filter
  adaptability, Nuclear Science and Techniques 33~(12) (2022) 159.
\newblock \href {https://doi.org/10.1007/s41365-022-01136-5}
  {\path{doi:10.1007/s41365-022-01136-5}}.

\bibitem{LIU20233359}
B.-Q. Liu, H.-R. Liu, L.~Chang, Y.-X. Cheng, Z.~Zuo, P.~Li, Discrimination of
  neutrons and gamma-rays in plastic scintillator based on spiking cortical
  model, Nuclear Engineering and Technology 55~(9) (2023) 3359--3366.
\newblock \href {https://doi.org/10.1016/j.net.2023.04.032}
  {\path{doi:10.1016/j.net.2023.04.032}}.

\bibitem{FABIAN2021164750}
X.~Fabian, G.~Baulieu, L.~Ducroux, O.~St{\'{e}}zowski, A.~Boujrad,
  E.~Cl{\'{e}}ment, S.~Coudert, G.~{de France}, N.~Erduran, S.~Ert{\"{u}}rk,
  V.~Gonz{\'{a}}lez, G.~Jaworski, J.~Nyberg, D.~Ralet, E.~Sanchis,
  R.~Wadsworth, {Artificial neural networks for neutron/{$\gamma$}
  discrimination in the neutron detectors of NEDA}, Nuclear Instruments and
  Methods in Physics Research Section A: Accelerators, Spectrometers, Detectors
  and Associated Equipment 986 (2021) 164750.
\newblock \href {https://doi.org/10.1016/j.nima.2020.164750}
  {\path{doi:10.1016/j.nima.2020.164750}}.

\bibitem{YOON20233925}
S.~Yoon, C.~Lee, H.~Seo, H.-D. Kim, {Improved fast neutron detection using
  CNN-based pulse shape discrimination}, Nuclear Engineering and Technology
  55~(11) (2023) 3925--3934.
\newblock \href {https://doi.org/10.1016/j.net.2023.07.007}
  {\path{doi:10.1016/j.net.2023.07.007}}.

\bibitem{Zhao_2023}
K.~Zhao, C.~Feng, S.~Wang, Z.~Shen, K.~Zhang, S.~Liu, {n/{$\gamma$}
  discrimination for CLYC detector using a one-dimensional Convolutional Neural
  Network}, Journal of Instrumentation 18~(01) (2023) P01021.
\newblock \href {https://doi.org/10.1088/1748-0221/18/01/P01021}
  {\path{doi:10.1088/1748-0221/18/01/P01021}}.

\bibitem{DOUCET2020161201}
E.~Doucet, T.~Brown, P.~Chowdhury, C.~Lister, C.~Morse, P.~Bender, A.~Rogers,
  {Machine learning n/{$\gamma$} discrimination in CLYC scintillators}, Nuclear
  Instruments and Methods in Physics Research Section A: Accelerators,
  Spectrometers, Detectors and Associated Equipment 954 (2020) 161201,
  symposium on Radiation Measurements and Applications XVII.
\newblock \href {https://doi.org/10.1016/j.nima.2018.09.036}
  {\path{doi:10.1016/j.nima.2018.09.036}}.

\bibitem{HACHEM20234057}
A.~Hachem, Y.~Moline, G.~Corre, B.~Ouni, M.~Trocme, A.~Elayeb, F.~Carrel,
  Labeling strategy to improve neutron/gamma discrimination with organic
  scintillator, Nuclear Engineering and Technology 55~(11) (2023) 4057--4065.
\newblock \href {https://doi.org/10.1016/j.net.2023.07.024}
  {\path{doi:10.1016/j.net.2023.07.024}}.

\bibitem{ZHANG2024111179}
S.~Zhang, Z.~Wei, P.~Zhang, Q.~Zhao, M.~Li, X.~Bai, K.~Wu, Y.~Nie, Y.~Ding,
  J.~Wang, Y.~Zhang, X.~Su, Z.~Yao, {Neutron-gamma discrimination with broaden
  the lower limit of energy threshold using BP neural network}, Applied
  Radiation and Isotopes 205 (2024) 111179.
\newblock \href {https://doi.org/10.1016/j.apradiso.2024.111179}
  {\path{doi:10.1016/j.apradiso.2024.111179}}.

\bibitem{10672537}
V.~H. Hai, N.~M. Dang, N.~T.~T. Phuc, H.~T.~K. Trang, T.~T.~H. Loan, P.~L.~H.
  Sang, M.~Nomachi, {Enhancing Neutron/Gamma Discrimination in the Low-Energy
  Region for EJ-276 Plastic Scintillation Detector Using Machine Learning},
  IEEE Transactions on Nuclear Science (2024) 1--1\href
  {https://doi.org/10.1109/TNS.2024.3456863}
  {\path{doi:10.1109/TNS.2024.3456863}}.

\bibitem{PAN2024103329}
Y.~Pan, P.~Gong, Z.~Hu, Z.~Wang, D.~Liang, C.~Zhou, X.~Zhu, X.~Tang, {Pulse
  pile-up recognition using multi-module DenseNet in neutron-gamma
  discrimination}, Nuclear Engineering and Technology (2024) 103329\href
  {https://doi.org/10.1016/j.net.2024.11.031}
  {\path{doi:10.1016/j.net.2024.11.031}}.

\bibitem{HAN2022166328}
J.~Han, J.~Zhu, Z.~Wang, G.~Qu, X.~Liu, W.~Lin, Z.~Xu, Y.~Huang, M.~Yan,
  X.~Zhang, L.~Chen, {Pulse characteristics of CLYC and piled-up
  neutron–gamma discrimination using a convolutional neural network}, Nuclear
  Instruments and Methods in Physics Research Section A: Accelerators,
  Spectrometers, Detectors and Associated Equipment 1028 (2022) 166328.
\newblock \href {https://doi.org/10.1016/j.nima.2022.166328}
  {\path{doi:10.1016/j.nima.2022.166328}}.

\bibitem{Peng_2022}
S.~Peng, Z.~Hua, Q.~Wu, J.~Han, S.~Qian, Z.~Wang, Q.~Wei, L.~Qin, L.~Ma,
  M.~Yan, R.~Song, {Piled-up neutron-gamma discrimination system for CLLB using
  convolutional neural network}, Journal of Instrumentation 17~(08) (2022)
  T08001.
\newblock \href {https://doi.org/10.1088/1748-0221/17/08/T08001}
  {\path{doi:10.1088/1748-0221/17/08/T08001}}.

\bibitem{MORALES2024745}
I.~R. Morales, M.~L. Crespo, M.~Bogovac, A.~Cicuttin, K.~Kanaki, S.~Carrato,
  {Gamma/neutron classification with SiPM CLYC detectors using frequency-domain
  analysis for embedded real-time applications}, Nuclear Engineering and
  Technology 56~(2) (2024) 745--752.
\newblock \href {https://doi.org/10.1016/j.net.2023.11.013}
  {\path{doi:10.1016/j.net.2023.11.013}}.

\bibitem{CLYCDataset}
M.~Argueta, I.~Ren{\'{e}}, {Gamma and neutron tagged dataset from CLYC SiPM
  detector} (2023).
\newblock \href {https://doi.org/10.5281/zenodo.8037239}
  {\path{doi:10.5281/zenodo.8037239}}.

\bibitem{DBLP:conf/icml/NairH10}
V.~Nair, G.~E. Hinton, Rectified linear units improve restricted boltzmann
  machines, in: J.~F{\"{u}}rnkranz, T.~Joachims (Eds.), Proceedings of the 27th
  International Conference on Machine Learning (ICML-10), June 21-24, 2010,
  Haifa, Israel, Omnipress, 2010, pp. 807--814.

\bibitem{ALHARBI2019205}
T.~Alharbi, Distance metrics for digital pulse-shape discrimination of
  scintillator detectors, Radiation Physics and Chemistry 156 (2019) 205--209.
\newblock \href {https://doi.org/10.1016/j.radphyschem.2018.11.014}
  {\path{doi:10.1016/j.radphyschem.2018.11.014}}.

\bibitem{detector-frontend}
{Thermal NEUTRON detector V12.7B30/SIP-E3-CLYC-X},
  \url{https://scionix.nl/wp-content/uploads/2017/03/V12.7B30_SIP-E3-CLYC-X.pdf},
  accessed: 2025-04-10.

\bibitem{DBLP:journals/corr/KingmaB14}
D.~P. Kingma, J.~Ba, Adam: {A} method for stochastic optimization, in:
  Y.~Bengio, Y.~LeCun (Eds.), 3rd International Conference on Learning
  Representations, {ICLR} 2015, San Diego, CA, USA, May 7-9, 2015, Conference
  Track Proceedings, 2015.

\bibitem{DBLP:conf/nips/PaszkeGMLBCKLGA19}
A.~Paszke, S.~Gross, F.~Massa, A.~Lerer, J.~Bradbury, G.~Chanan, T.~Killeen,
  Z.~Lin, N.~Gimelshein, L.~Antiga, A.~Desmaison, A.~K{\"{o}}pf, E.~Z. Yang,
  Z.~DeVito, M.~Raison, A.~Tejani, S.~Chilamkurthy, B.~Steiner, L.~Fang,
  J.~Bai, S.~Chintala, Pytorch: An imperative style, high-performance deep
  learning library, in: H.~M. Wallach, H.~Larochelle, A.~Beygelzimer,
  F.~d'Alch{\'{e}}{-}Buc, E.~B. Fox, R.~Garnett (Eds.), Annual Conference on
  Neural Information Processing Systems 2019, NeurIPS 2019, December 8-14,
  2019, Vancouver, BC, Canada, 2019, pp. 8024--8035.

\bibitem{CLYC-scint}
{Cesium Yttrium Chloride (CLYC:Ce) for Dual Neutron {\&} Gamma Radiation
  Detection}, \url{https://www.symmic.net/docs/CLYC.pdf}, accessed: 2025-04-10.

\bibitem{MICROC-SiPM}
{C-Series SiPM Sensors},
  \url{https://www.onsemi.cn/download/data-sheet/pdf/microc-series-d.pdf},
  accessed: 2025-04-10.

\bibitem{NGMethods}
P.~Li, H.~Liu, Dataset for neutron and gamma-ray pulse shape discrimination:
  radiation pulse signals and discrimination methodologies (2024).
\newblock \href {https://doi.org/10.5281/zenodo.10947029}
  {\path{doi:10.5281/zenodo.10947029}}.

\bibitem{TOFDataset}
P.~Maedgen, B.~Wellons, S.~Prasad, J.~Tao, {Improving Pulse Shape
  Discrimination in Organic Scintillation Detectors by Understanding Underlying
  Data Structure (dataset)},
  \url{https://github.com/NeutronNeutrinoSensing/PSDwithML}, accessed:
  2025-04-10.

\end{thebibliography}





\end{document}